\documentclass[useAMS,usenatbib,usegraphicx,letterpaper]{mn2e}
\usepackage{times,amssymb,hyperref,subfigure,epsfig,aas_macros}

\usepackage[totalwidth=480pt,totalheight=680pt,layoutvoffset=0.5cm]{geometry}

\begin{document}

\title[Kuiper belts around super-Earth hosts]{Kuiper belt structure around nearby
  super-Earth host stars}

\author[Grant M. Kennedy et al]{Grant M. Kennedy\thanks{Email:
    \href{mailto:gkennedy@ast.cam.ac.uk}{gkennedy@ast.cam.ac.uk}}$^1$, Luca
  Matr\`a$^{1,2}$, Maxime Marmier$^3$, Jane S. Greaves$^4$, \newauthor Mark C. Wyatt$^1$,
  Geoffrey Bryden$^5$, Wayne Holland$^{6,7}$, Christophe Lovis$^3$, \newauthor Brenda
  C. Matthews$^{8,9}$, Francesco Pepe$^3$, Bruce Sibthorpe$^{10}$, St\'ephane Udry$^3$ \\
  $^1$ Institute of Astronomy, University of Cambridge, Madingley Road, Cambridge CB3
  0HA, UK \\
  $^2$ European Southern Observatory, Alonso de C\'ordova 3107, Vitacura, Santiago,
  Chile\\
  $^3$ D\'epartement d'Astronomie de l'Universit\'e de Gen\`eve, 51 ch. des Maillettes -
  Observatoire de Sauverny, CH-1290 Versoix, Switzerland \\
  $^4$ School of Physics and Astronomy, University of St Andrews, North Haugh, St
  Andrews, Fife KY16 9SS, UK\\
  $^5$ Jet Propulsion Laboratory, California Institute of Technology, 4800 Oak Grove
  Drive, Pasadena, CA 91109, USA \\
  $^6$ UK Astronomy Technology Center, Royal Observatory, Blackford Hill, Edinburgh EH9
  3HJ, UK\\
  $^7$ Institute for Astronomy, University of Edinburgh, Royal Observatory, Blackford
  Hill,
  Edinburgh EH9 3HJ, UK \\
  $^8$ National Research Council of Canada, 5071 West Saanich Road, Victoria, BC, Canada V9E 2E7\\
  $^9$ University of Victoria, Finnerty Road, Victoria, BC, V8W 3P6, Canada\\
  $^{10}$ SRON Netherlands Institute for Space Research, NL-9747 AD Groningen, The
  Netherlands \\
}
\maketitle

\begin{abstract}
  We present new observations of the Kuiper belt analogues around HD~38858 and HD~20794,
  hosts of super-Earth mass planets within 1 au. As two of the four nearby G-type stars
  (with HD~69830 and 61~Vir) that form the basis of a possible correlation between
  low-mass planets and debris disc brightness, these systems are of particular
  interest. The disc around HD~38858 is well resolved with \emph{Herschel} and we
  constrain the disc geometry and radial structure. We also present a probable JCMT
  sub-mm continuum detection of the disc and a CO J=2-1 upper limit. The disc around
  HD~20794 is much fainter and appears marginally resolved with \emph{Herschel}, and is
  constrained to be less extended than the discs around 61~Vir and HD~38858. We also set
  limits on the radial location of hot dust recently detected around HD~20794 with
  near-IR interferometry. We present HARPS upper limits on unseen planets in these four
  systems, ruling out additional super-Earths within a few au, and Saturn-mass planets
  within 10 au. We consider the disc structure in the three systems with Kuiper belt
  analogues (HD~69830 has only a warm dust detection), concluding that 61~Vir and
  HD~38858 have greater radial disc extent than HD~20794. We speculate that the greater
  width is related to the greater minimum planet masses (10-20 $M_\oplus$ vs. 3-5
  $M_\oplus$), arising from an eccentric planetesimal population analogous to the Solar
  System's scattered disc. We discuss alternative scenarios and possible means to
  distinguish among them.
\end{abstract}

\begin{keywords}
  planetary systems: formation --- planet-disc interactions --- circumstellar matter ---
  stars: individual: HD~20794 --- stars: individual: 61~Vir --- stars: individual:
  HD~69830 --- stars: individual: HD~38858
\end{keywords}

\section{Introduction}\label{s:intro}

The first evidence for the existence of a planetary system other than our own was
arguably the image of the debris disc around $\beta$ Pictoris
\citep{1984Sci...226.1421S}. With the contemporaneous discovery of many other examples of
the ``Vega phenomenon'' with the IRAS satellite \citep{1984ApJ...278L..23A}, it was soon
realised that the dust in these systems was the product of collisions between larger
planetesimals \citep{1993prpl.conf.1253B}, the same building blocks thought to build the
planets now seen around hundreds of stars.

Despite a clear conceptual connection between planets and debris discs, both being a
product of some degree of planet formation, there has until recently been surprisingly
little evidence of statistically significant correlations between the two
\citep[e.g.][]{2009ApJ...705.1226B,2009ApJ...700L..73K,2012MNRAS.426...91K,2012ApJ...757....7M}. However,
with many nearby stars observed with both the \emph{Spitzer} \citep{2004ApJS..154....1W}
and \emph{Herschel}\footnote{Herschel is an ESA space observatory with science
  instruments provided by European-led Principal Investigator consortia and with
  important participation from NASA.} \citep{2010A&A...518L...1P} observatories, and the
continual increase in the sensitivity of radial velocity surveys for planets, the first
correlations have begun to emerge
\citep{2012MNRAS.424.1206W,2014A&A...565A..15M}. \citet{2012MNRAS.424.1206W} reported
that of 6 out of the 60 nearest G-type stars with close-in low-mass planets and no known
planets above a Saturn-mass, 4 have relatively bright debris discs (i.e. at least an
order of magnitude brighter than our Edgeworth-Kuiper belt; HD~20794, 61~Vir, HD~69830,
HD~38858). This high detection rate is in contrast with a detection rate of $\sim$15\%
around normal Sun-like main-sequence stars \citep{2008ApJ...674.1086T}, which is
comparable with that for stars with more massive planets, suggesting there is some kind
of link between low-mass planets and the brightness of the debris in those systems. A
decreased ability of low-mass planets to eject planetesimals compared to gas giants is a
possible explanation \citep[e.g.][]{2009MNRAS.393.1219P,2012MNRAS.424.1206W}.

With only a handful of low-mass planet + disc systems, the cause of such a correlation
remains unclear, so characterisation of the four G-type systems that contribute to this
correlation is a first step towards an understanding. Of those four, resolved imaging of
61 Vir was presented and discussed by \citet{2012MNRAS.424.1206W}, and HD~69830, which
only hosts warm dust that was not detected by \emph{Herschel}
\citep{2014A&A...565A..15M}, has been discussed in numerous studies
\citep{2005ApJ...626.1061B,2007ApJ...658..584L,2009A&A...503..265S,2011ApJ...743...85B}.
Here, we present new \emph{Herschel} PACS observations of the remaining two systems;
HD~38858 and HD~20794. The former hosts a bright and well resolved debris disc, so was
also observed with SCUBA-2 and RxA3 on the JCMT. The latter hosts one of the faintest
debris discs known, and at only 6 pc and with three low-mass planets is clearly a system
of high interest. In what follows we present and describe the observations, derive
constraints on the disc spatial structure, present upper limits on undiscovered planets
using radial velocity (RV) data, and discuss the disc structure and possible relations
with the planets.

\section{The HD~38858 and HD~20794 systems}\label{s:sys}

We begin by describing the previously known stellar and planetary properties of our
systems. HD~38858 (HIP~27435) is a nearby (15.2 pc) Sun-like (G4V) star.  Stellar age
estimates vary from as young as 200 Myr \citep{2011A&A...530A.138C}, to as old as 2.32 to
8.08 Gyr \citep{2007ApJS..168..297T}, and 9.3 Gyr \citep{2010A&A...512L...5S}. The star
is known to host a cool debris disc, which was first detected with \emph{Spitzer}
\citep{2006ApJ...652.1674B}. As noted by \citet{2012AJ....144...45K}, the disc was found
to be marginally resolved at 70 $\mu$m with \emph{Spitzer} with an inclination of
48$^\circ$, position angle of 56$^\circ$, and a radius 135 au (9\arcsec, no uncertainties
given). The star was observed with the Precision Integrated Optics Near Infrared
ExpeRiment \citep[PIONIER,][]{2011A&A...535A..67L} in search of a resolved excess, and
the 3$\sigma$ upper limit on the disc/star flux ratio in the H-band is 1\%
\citep{2014A&A...570A.128E}. The star has been reported to host a super-Earth mass planet
\citep{2011arXiv1109.2497M}, for which we provide an update on the orbit below (Marmier
et al. in preparation).

HD~20794 (HIP~15510) is a nearby (6 pc) Sun-like (G8V) star. Stellar age estimates vary
from as young as 3.9 Gyr \citep{2011A&A...530A.138C}, through intermediate ages of 5.2 to
6.2 Gyr \citep{2012AJ....143..135V} and up to 11.3 Gyr \citep{2009A&A...501..941H}. The
star has been found to host two robust low-mass planets, with a possible third planet
\citep{2011A&A...534A..58P}. The star was subsequently found to host a debris disc using
archival \emph{Spitzer} data, as described by \citet{2012MNRAS.424.1206W}, who also
reported a detection with \emph{Herschel} that we elaborate on below. This star was also
observed with PIONIER, resulting in a significant $H$-band disc/star flux ratio of $1.64
\pm 0.37$ \citep{2014A&A...570A.128E}.

\section{Debris Disc Observations}\label{s:obs}

Here we present new observations of HD~38858 with several different instruments. It was
observed by \emph{Herschel} \citep{2010A&A...518L...1P} under the auspices of the Search
for Kuiper-belts ARound Planet-host Stars (SKARPS) survey
\citep[e.g.][]{2013MNRAS.436..898K}, and by the new Submillimetre Common User Bolometer
Array (SCUBA-2) as part of the SCUBA-2 Observations of Nearby Stars (SONS) James Clerk
Maxwell Telescope (JCMT) Legacy survey \citep{2013MNRAS.435.1037P}. Finally, HD~38858 has
also been observed in search of CO J=2-1 emission using the RxA3 receiver on the JCMT.

We also present a detailed analysis of observations of HD~20794 with the \emph{Herschel}
PACS instrument, observed as part of the Disc Emission via a Bias-free Reconnaissance in
the Infrared/Submillimetre (DEBRIS) survey. The disc around this star is very faint
compared to HD~38858, so it was not selected for the any of our JCMT programmes due to
the high probability of a non-detection.

\subsection{HD~38858}\label{ss:38858obs}

\subsubsection{Herschel PACS}

\begin{table}
  \caption{\emph{Herschel} PACS observations of HD~38858 and HD~20794. All PACS
    observations use either the 70 or 100 $\mu$m bands, and always include the 160 $\mu$m
    band.}\label{tab:obs}
  \begin{tabular}{lllll}
    \hline
    Target & ObsIds & Date & $\lambda$ ($\mu$m) & Duration (s) \\
    \hline
    HD~20794 & 1342216456/7 & 20 Mar 2011 & 100 & $2 \times 1686$ \\
    HD~20794 & 1342234096/7 & 14 Dec 2011 & 70 & $2 \times 445$ \\
    HD~38858 & 1342242537/8 & 28 Mar 2012 & 70 & $2 \times 1686$ \\
    \hline
  \end{tabular}  
\end{table}

\begin{figure*}
  \begin{center}
    \hspace{-0.25cm} \includegraphics[width=0.48\textwidth]{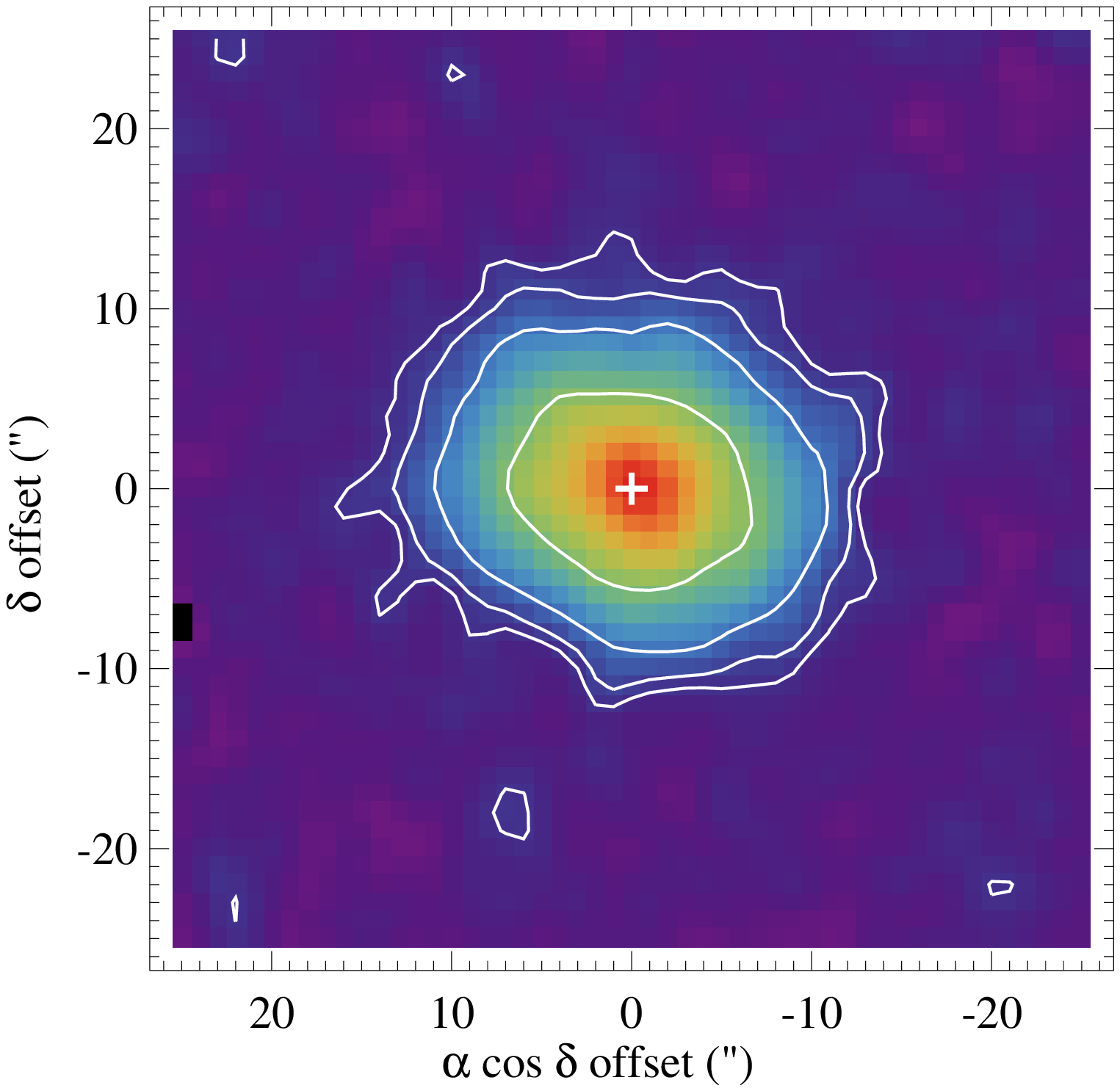}
    \hspace{-2cm} \includegraphics[width=0.48\textwidth]{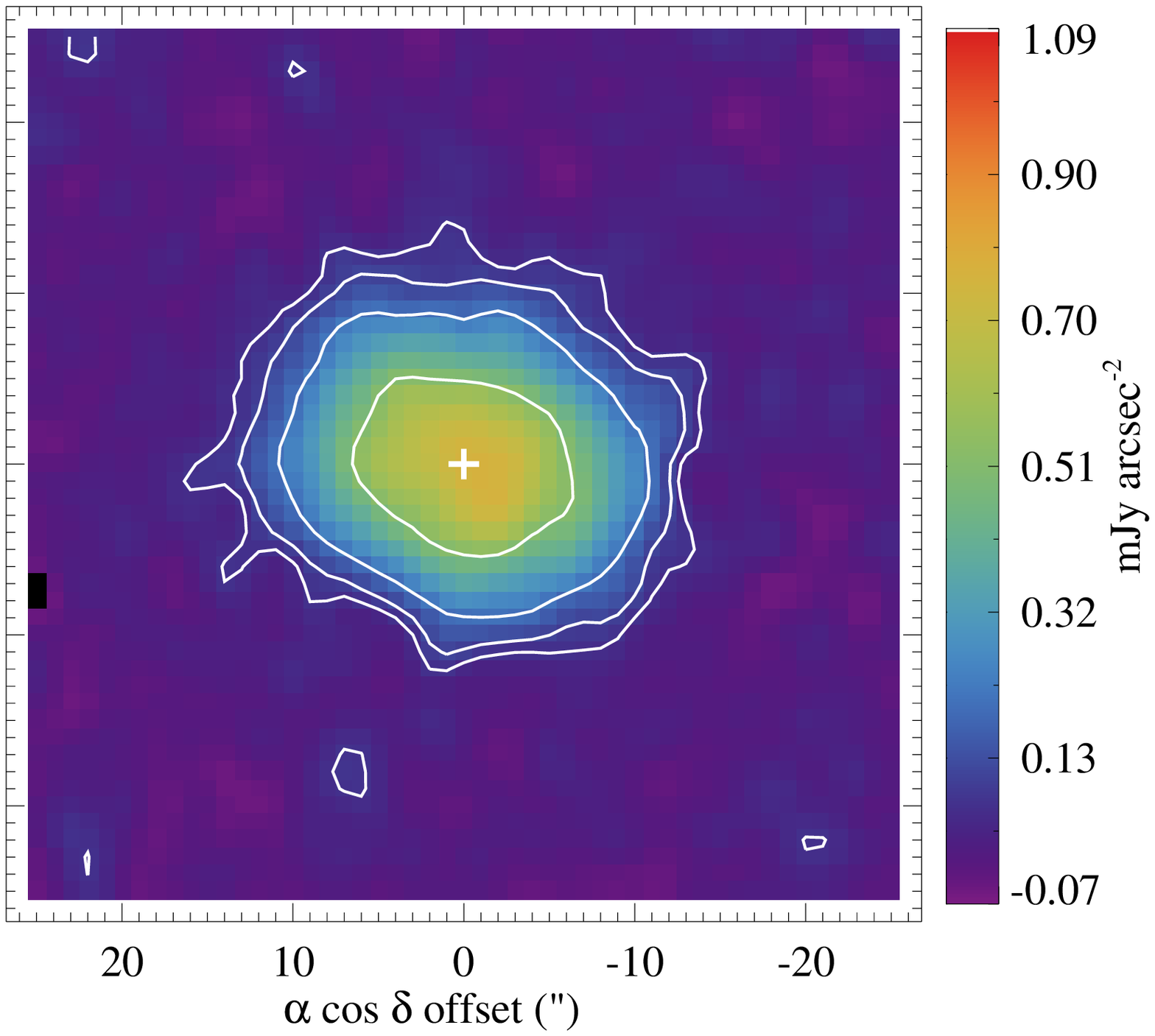} \\
    \caption{\emph{Herschel} PACS 70 $\mu$m images of HD~38858. The left panel shows the
      original image and the star has been subtracted from the right panel. The star
      location is indicated by the white cross. Contours are at 3, 5, 10, and 25 times
      the pixel RMS level of $2 \times 10^{-2}$ mJy arcsec$^{-2}$ (70 $\mu$m pixels are 1
      arcsec$^2$). The PACS beam size at 70 $\mu$m is about 6\arcsec, so the disc is well
      resolved.}\label{fig:38858ims}
  \end{center}
\end{figure*}

HD~38858 was observed with the \emph{Herschel} Photodetector Array Camera and
Spectrometer \citep[PACS,][]{2010A&A...518L...2P} instrument at 70 and 160 $\mu$m as part
of the SKARPS open time programme (see Table \ref{tab:obs}). The 70 $\mu$m image of
HD~38858 is shown in Fig. \ref{fig:38858ims}. The left panel shows the original image and
the stellar emission has been subtracted from the right panel. Most of the 70 $\mu$m
emission clearly originates in the disc, and the full-width at half-maximum (FWHM) PACS
beam size at 70 $\mu$m is about 6\arcsec~so the disc is clearly resolved. The disc is
approximately 15\arcsec~across, so the system distance of 15.2 pc implies a disc diameter
of around 225 au, similar to the size estimated from \emph{Spitzer}
\citep{2012AJ....144...45K}. We return to the disc properties and geometry in section
\ref{ss:38858mods}.

Measuring the brightness of the star+disc in the PACS image is fairly straightforward for
HD~38858. The disc is well resolved so we use a 20\arcsec~radius aperture that is centred
on the peak emission location (and about 3\arcsec~from the expected stellar position, so
consistent with the pointing uncertainty). After aperture correction we find a flux
density of $190 \pm 6$ mJy at 70 $\mu$m, and of $141 \pm 20$ mJy at 160 $\mu$m. The
uncertainty is estimated by the mean value of a number of 20\arcsec~apertures placed
randomly in high coverage regions in each map, and is added in quadrature to the
calibration uncertainty (which is included above, and dominates the 70 $\mu$m
uncertainty). The uncertainty is several times higher at 160 $\mu$m than expected simply
from the PACS integration time due to a high Galactic background level. This high
background level is expected in the Orion Complex, and HD~38858 was excluded from the
DEBRIS survey based on such an expectation \citep{2010MNRAS.403.1089P}. These
measurements are in excellent agreement with a 70 $\mu$m \emph{Spitzer} measurement of
$190 \pm 9$ mJy \citep{2009ApJ...705...89L}.

\subsubsection{JCMT SCUBA-2 continuum imaging}\label{sss:s2}

\begin{figure*}
  \begin{center}
    \hspace{-0.25cm} \includegraphics[width=0.48\textwidth]{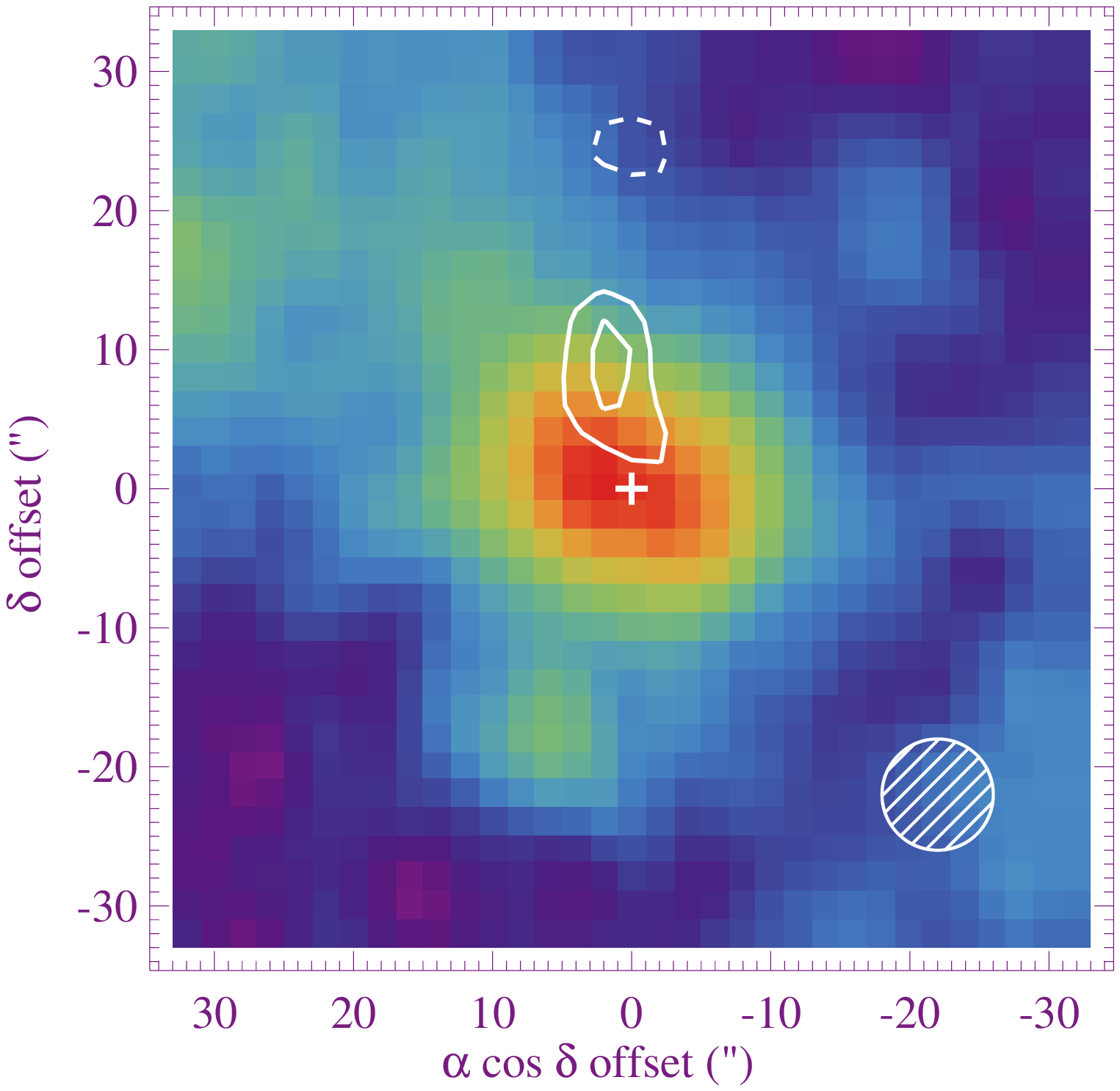}
    \hspace{-2cm} \includegraphics[width=0.48\textwidth]{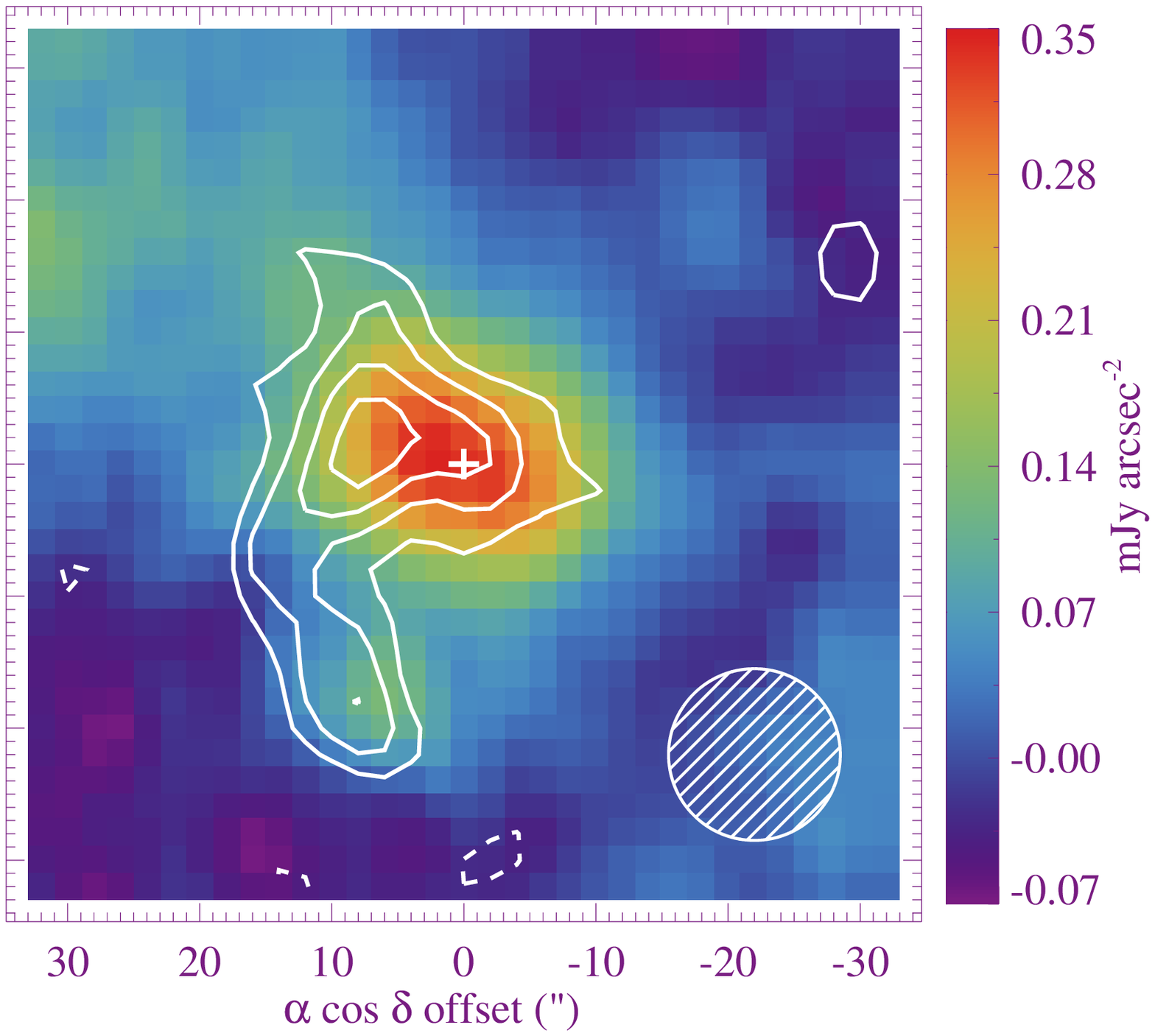}
    \caption{SCUBA-2 images of HD~38858 at 450 (left panel) and 850 $\mu$m (right
      panel). The contours show the SCUBA-2 images with 3, 4, 5 and 6$\sigma$ contours
      (solid are positive, dashed are negative), and the rainbow coloured background
      image in both panels shows the 160 $\mu$m PACS image. The star location is
      indicated by the white cross. As indicated by the circles in the lower right of
      each panel, the SCUBA-2 beam FWHM is 8 and 13\arcsec~at 450 and 850
      $\mu$m.}\label{fig:38858s2}
  \end{center}
\end{figure*}

As a star with a disc bright enough for a possible detection at sub-mm wavelengths,
HD~38858 was included in the SONS survey, which uses SCUBA-2 \citep{2013MNRAS.430.2513H}
to detect and image discs at 450 and 850~$\mu$m. HD~38858 was observed for a total of 10
hours, made up of 20 half-hour integrations. We used the constant speed DAISY pattern,
which provides uniform exposure time coverage in the central 3 arcmin-diameter region of
a field \citep{2013MNRAS.430.2513H}.

The SCUBA-2 data were reduced using the Dynamic Iterative Map-Maker within the STARLINK
SMURF package \citep{2013MNRAS.430.2545C} called from the ORAC-DR automated pipeline
\citep{2008AN....329..295C}. The map maker used a configuration file optimized for known
position, compact sources. It adopts the technique of ``zero-masking'' in which the map
is constrained to a mean value of zero (in this case outside a radius of 60\arcsec~from
the center of the field), for all but the final interation of the map maker
\citep{2013MNRAS.430.2545C}. The technique not only helps convergence in the iterative
part of the map-making process but suppresses the large-scale ripples that can produce
ringing artefacts. The data are also high-pass filtered at 1 Hz, corresponding to a
spatial cut-off of $\sim$150\arcsec~for a typical DAISY scanning speed of 155\arcsec
/s. The filtering removes residual low-frequency (large spatial scale) noise and, along
with the ``zero-masking'' technique, produces flat and uniform final images largely
devoid of gradients and artefacts \citep{2013MNRAS.430.2545C}.

To account for the attenuation of the signal as a result of the time series filtering,
the pipeline re-makes each map with a fake 10 Jy Gaussian added to the raw data, but
offset from the nominal map centre by 30\arcsec~to avoid contamination with any detected
source. The amplitude of the Gaussian in the output map gives the signal attenuation, and
this correction is applied along with the flux conversion factor derived from the
calibrator observations. The final images were made by coadding the 20 maps using
inverse-variance weighting, re-gridded with 1\arcsec~pixels at both wavelengths. The
final images at both wavelengths have been smoothed with a 7\arcsec~FWHM Gaussian to
improve the signal-to-noise ratio. The FWHMs of the primary beam are 7.9\arcsec~and
13.0\arcsec~at 450 and 850 $\mu$m, respectively.

Fig. \ref{fig:38858s2} shows the SCUBA-2 images overlaid as contours on the PACS 160
$\mu$m data. At 450 $\mu$m a 4$\sigma$ significant peak is seen about 10\arcsec~to the
North of HD~38858. This detection is consistent with being unresolved, and has a flux
density of $49 \pm 11$ mJy. The relatively low S/N means that the detected flux may
actually originate closer to HD~38858 than 10\arcsec, but given the known high background
level it is likely that this detection is not associated with HD~38858 and we do not
consider it further.

At 850 $\mu$m the SCUBA-2 data show a clear detection. The peak location (6$\sigma$) is
offset from the star location by about 10\arcsec~to the East, and the centre of the
5$\sigma$ contour offset by about 5\arcsec. A background source is seen to the South, and
probably corresponds to emission detected at 160 $\mu$m at about the same location with
PACS. The co-location of these sources suggests that the accuracy of the JCMT pointing
for these observations is within a few seconds of arc, and splitting the SCUBA-2 data in
half yields two images with similar structure. More broadly, we find no systematic
pointing errors in SONS observations, for example the pre- and post-observation pointing
calibrations are within about 3\arcsec~across the entire SONS survey.  The 850 $\mu$m
flux density at the peak position is $4.9 \pm 0.7$ mJy/beam, and at the stellar position
$3.7 \pm 0.7$ mJy/beam. The integrated flux within 30\arcsec radius of the peak location
is $11.5 \pm 1.3$ mJy, the larger flux indicating that the emission covers a few JCMT
resolution elements. The 3-5$\sigma$ contours cover the location of HD~38858, so the
image could include both Orion Complex background emission to the East of the star and
disc emission centred on the star. If the contamination is not severe, the disc flux lies
between 3.7 and 11.5 mJy.

An alternative possibility is extra-galactic confusion; given about $10^3$ galaxies
degree$^{-2}$ brighter than 5 mJy \citep[e.g.][]{1999ApJ...512L..87B}, there is a 2\%
chance that a source as bright as the peak would appear within an aperture 15\arcsec~in
radius centred on HD~38858, and a much smaller chance that such a source would be well
centred on the star. Thus, the star-centred emission could arise from a galaxy with
$\sim$0.2\% probability, or the offset peak could arise from a galaxy with 2\%
probability. In the latter case the star-centred emission still requires an explanation,
and adding a second galaxy lowers the probability significantly. Thus, in either case it
is unlikely that the emission located at the stellar position originates from a
background galaxy.
% Given that the 70 $\mu$m PACS image shows no evidence for such a source the fact that
% HD~38858 was chosen to be observed based on the large 70 $\mu$m flux does not enhance
% this likelihood.

Given the small chance of contamination from background galaxies, and that some
contribution from the Orion Complex is likely, we consider the SCUBA-2 emission a
probable detection of emission associated with HD~38858. Our favoured interpretation is
that because the 3-5$\sigma$ contours cover the stellar location, 850$\mu$m emission from
the disc is indeed detected with a flux of at least 3.5 mJy (i.e. $>$5$\times$ the point
source uncertainty) and that the emission is shifted to the East by extended background
contamination that also includes the source to the South. We illustrate this
interpretation further after constructing models of the HD~38858 disc below in section
\ref{sss:s2mod}, where we find an 850 $\mu$m disc flux of $7.5 \pm 2$ mJy.

\subsubsection{JCMT CO J=2-1 line emission}\label{sss:co}

\begin{figure}
  \begin{center}
    \hspace{-0.5cm} \includegraphics[width=0.48\textwidth]{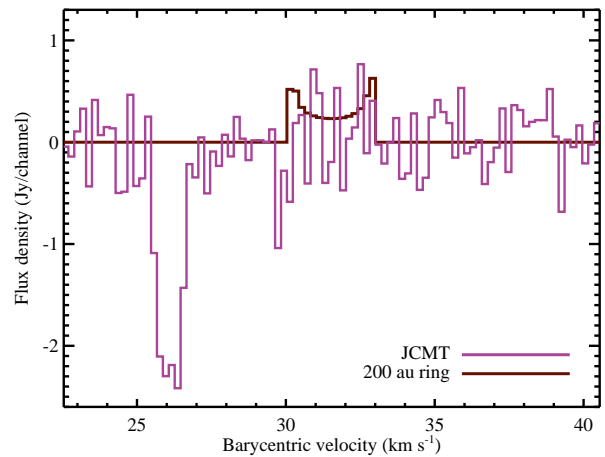}
%    \hspace{0.2cm}
%    \includegraphics[width=0.48\textwidth]{figs/co-r.eps}
    \caption{JCMT RxA3 spectrum for HD~38858. The velocity of HD~38858 is $31.54 \pm
      0.04$ km $s^{-1}$. The dark red line shows the velocity profile for a ring of material
      at 200 au at the detection limit of 1 Jy km s$^{-1}$.
%      Right panel: Significance analysis showing the measured line intensity
%    compared to the 3$\sigma$ uncertainty derived by randomising the spectrum. Around
%      180 au the line intensity is 4$\sigma$ above the estimated noise level.
    }\label{fig:38858co}
  \end{center}
\end{figure}

The RxA3 receiver on the JCMT was used to search for CO J=2-1 (230.538 GHz) emission
towards HD 38858. The grid-chop mode was used to stare at the target while beam-switching
on the sky.  Standard calibration was done using hot, cold and sky loads, and inspection
of spectra of standard sources. The co-added spectrum consists of 27 observations of 20
minutes duration each, over 5 nights between 12 Nov 2013 and 24 Dec 2013. The correction
$\eta$(main-beam) ($T_{\rm mb} = T_{\rm A}^*/\eta_{\rm mb})$ at the time was between 0.5
and 0.6 from measurements on Mars. The 22 individual outputs of the ORACDR pipeline were
co-added using the 'wcsmosaic' task in KAPPA. The output data have units of $T_{\rm A}^*$
in K versus heliocentric velocity (radio definition) in km s$^{-1}$, and are converted to
Jy with multiplication by a factor of 27.4 Jy K$^{-1}$. The spectral 'cube' (1x1x2048
points) was exported to Starlink SPLAT for further reduction. The spectrum was rebinned
by a factor of 5 to a 0.2 km s$^{-1}$ channel width, and a low-order polynomial fit (that
excluded the absorption feature described below) subtracted.

% This consisted of removal of a 3 mK residual baseline; removal via a Gaussian fit of a
% negative feature of peak $T_{\rm A}^*$ = 90 mK (2.65 Jy) and FWHM = 0.9 km s$^{-1}$ at
% 25.8 km s$^{-1}$ (possibly due to chopping onto an unrelated interstellar cloud). The
% data were then binned by 5 spectrally to a 0.2 km s$^{-1}$ channel width, and a second
% iteration of the baseline of 1 mK was removed.

The RxA3 spectrum is shown in Fig. \ref{fig:38858co}, in Jy channel$^{-1}$. A spectral
line associated with the star should be centred around the heliocentric stellar radial
velocity of 31.54 km s$^{-1}$ \citep{2002ApJS..141..503N}. To guide the eye the dark red
line shows an example line profile with an integrated flux of 1 Jy km s$^{-1}$,
calculated as the histogram of velocities for material uniformly distributed around a
circular narrow ring at 200 au, with the inclination derived from the 70 $\mu$m PACS
image. There is no clear line emission at the expected velocity of HD~38858, but a large
negative feature at 26 km s$^{-1}$, which likely originates from the Orion Complex
background, where CO J=1-0 emission is seen at 23.5 to 30 km s$^{-1}$
\citep[heliocentric,][]{2005A&A...430..523W}. This dip in the spectrum makes estimation
of the baseline difficult; the flux in the range 27-30 km s$^{-1}$ may suffer from
absorption and thus the flux near the velocity of HD~38858 could be higher than estimated
in Fig. \ref{fig:38858co}. Future observations that find the origin of this line and
subtract it will help resolve this issue.

We estimate the uncertainty on the measurement by fitting line profiles like that shown
in Fig. \ref{fig:38858co} for a range of velocity widths. We first randomized the bins in
the spectrum multiple times, fitting lines of a range of widths each time, resulting in a
standard deviation of about 1 Jy km s$^{-1}$ for radii from 50-300 au. We repeated this
exercise, but at a range of locations in the non-randomized spectrum, finding similar
results. We revisit the implications of this upper limit in terms of the gas mass in
section \ref{ss:gas}.

\subsection{HD~20794}\label{ss:20794obs}

We now turn to the observations of HD~20794. The disc around this star is much fainter
than that around HD~38858 so was only observed with \emph{Herschel} PACS (see Table
\ref{tab:obs}). HD~20794 was first observed at 100 $\mu$m and 160$\mu$m, and following
the possible resolution of the disc at low S/N, was re-observed at 70 and 160
$\mu$m.

\begin{figure*}
  \begin{center}
    \vspace{-0.25cm}
    \hspace{-0.25cm} \includegraphics[width=0.48\textwidth]{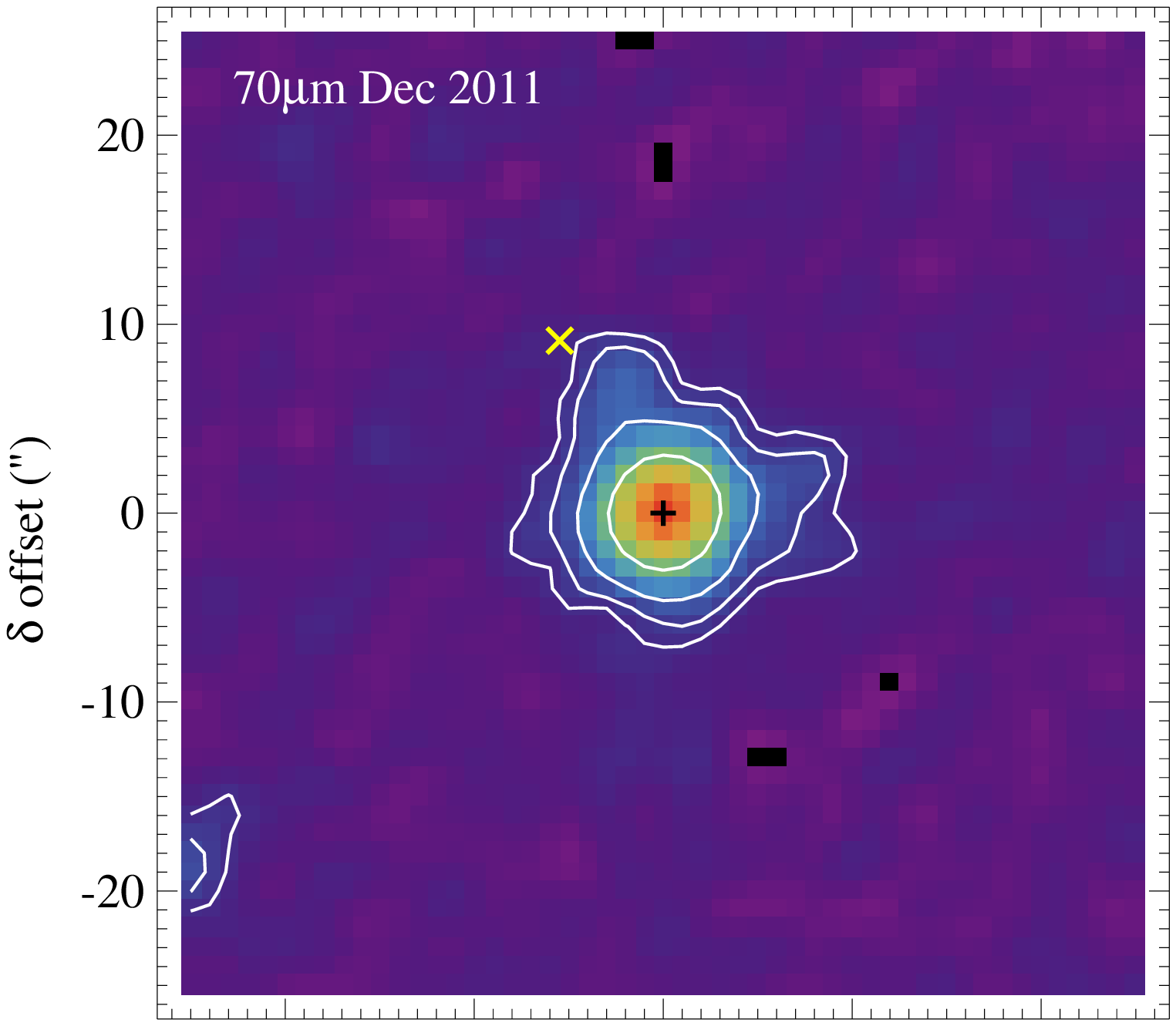}
    \hspace{-2.5cm} \includegraphics[width=0.48\textwidth]{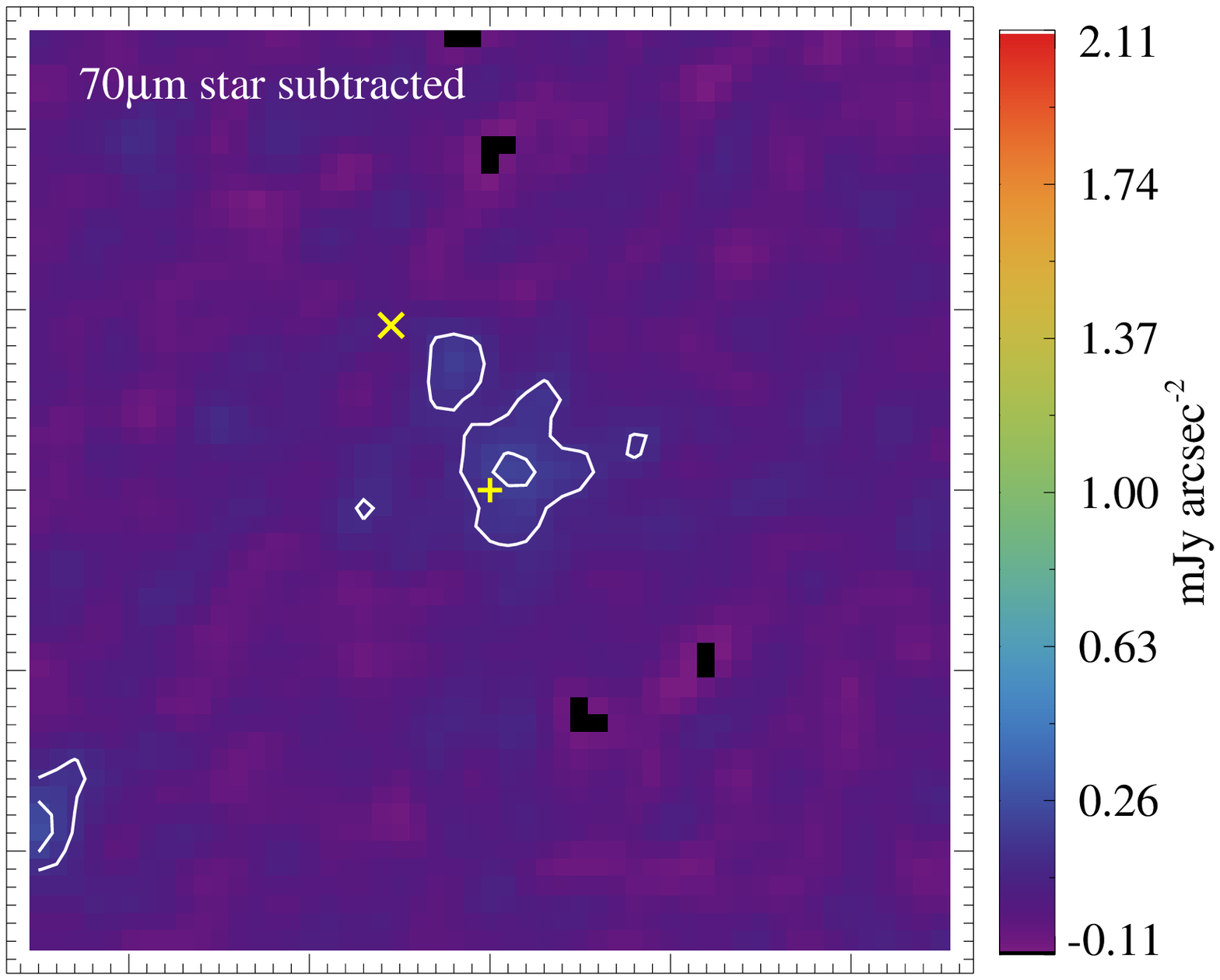} \\
    \vspace{-1.2cm}
    \hspace{-0.25cm} \includegraphics[width=0.48\textwidth]{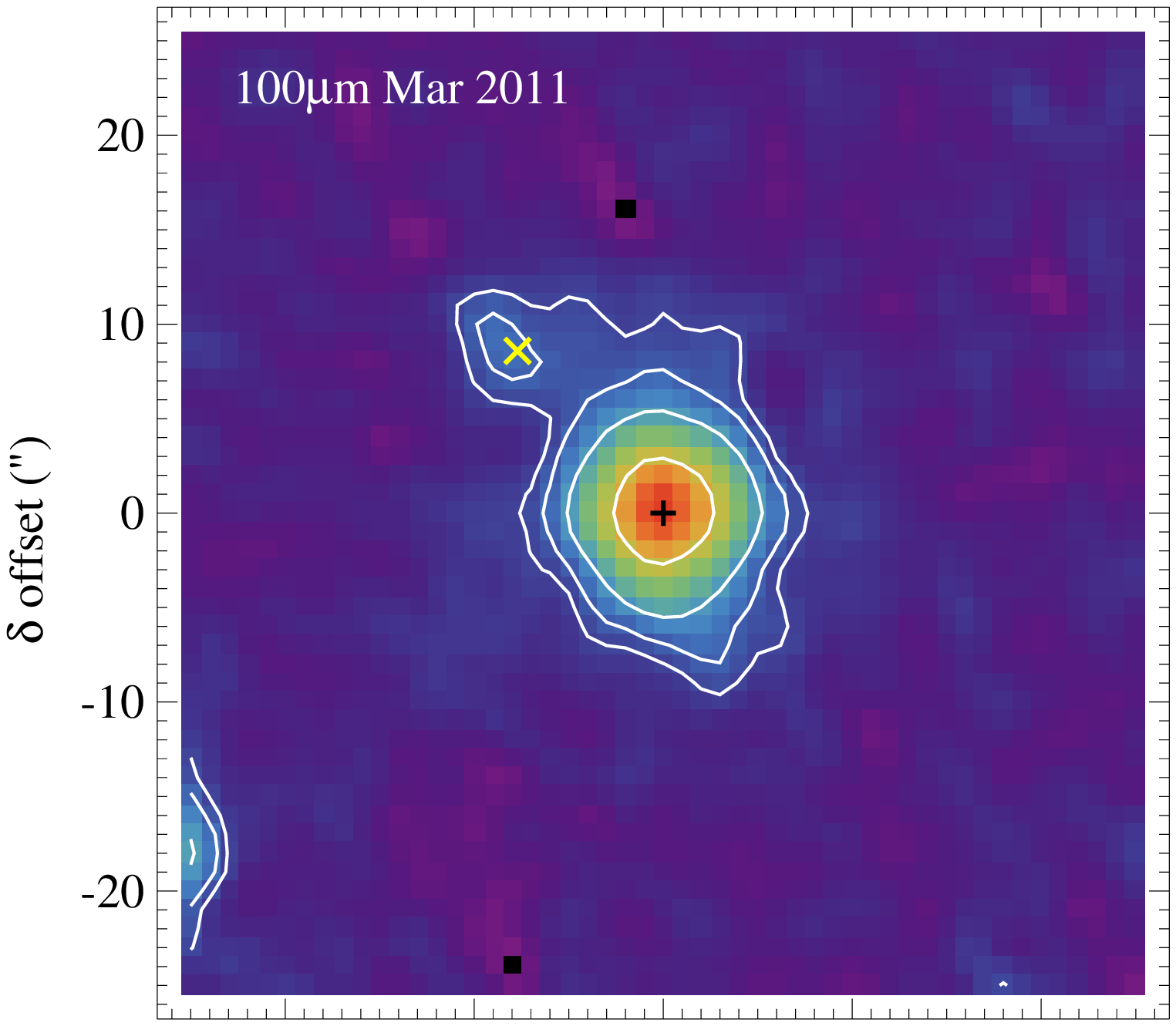}
    \hspace{-2.5cm} \includegraphics[width=0.48\textwidth]{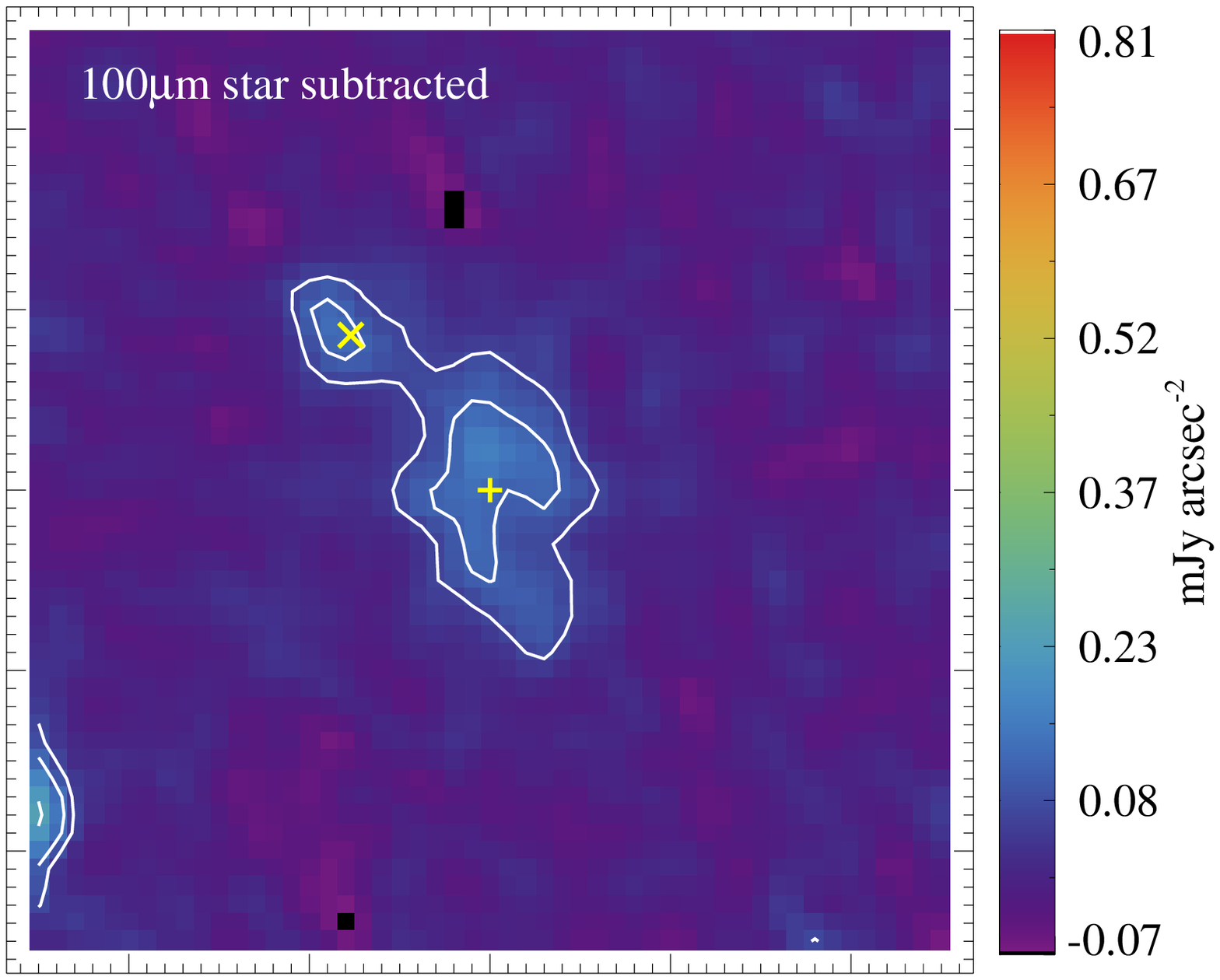} \\
    \vspace{-1.2cm}
    \hspace{-0.25cm} \includegraphics[width=0.48\textwidth]{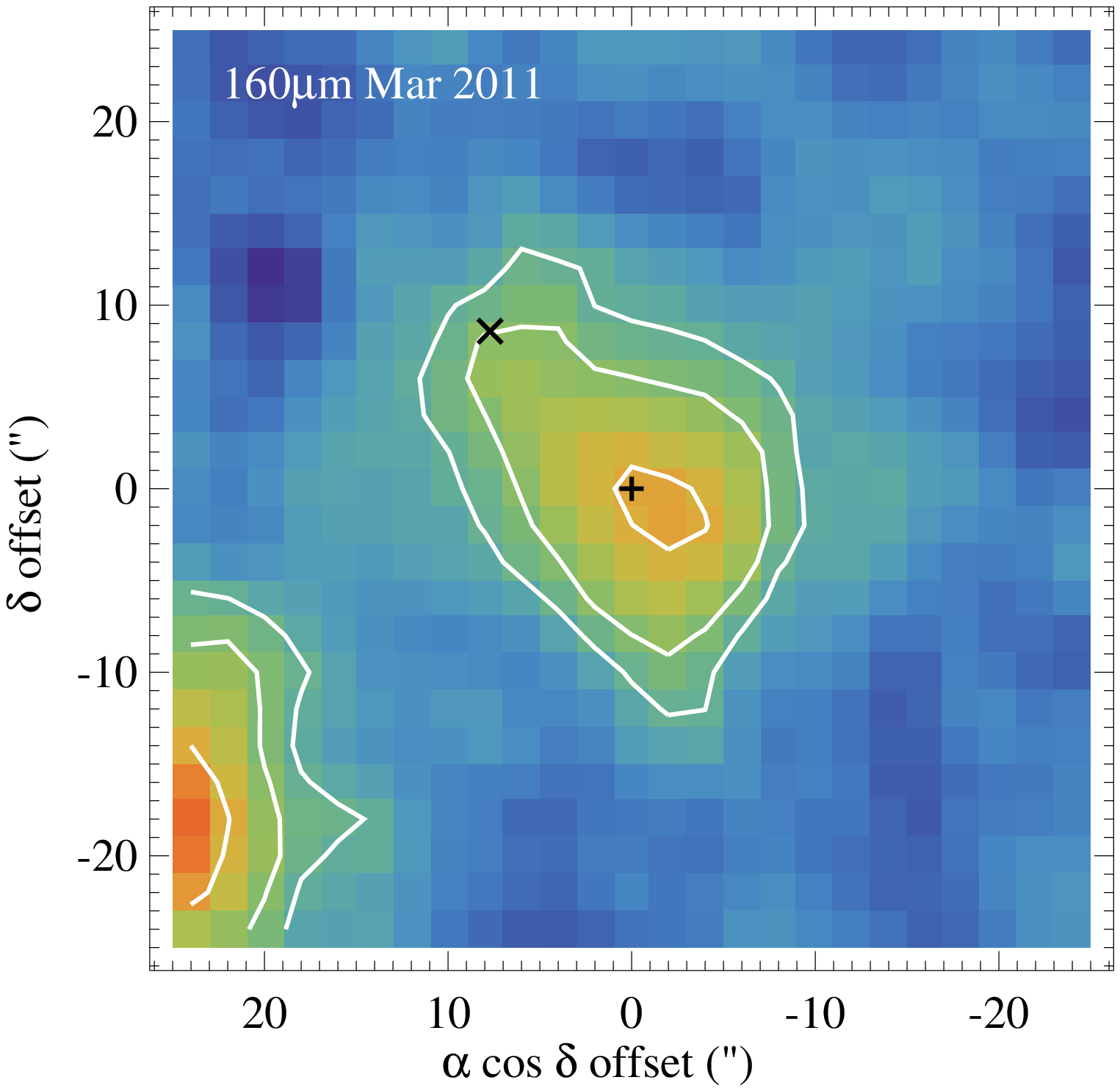}
    \hspace{-2.5cm} \includegraphics[width=0.48\textwidth]{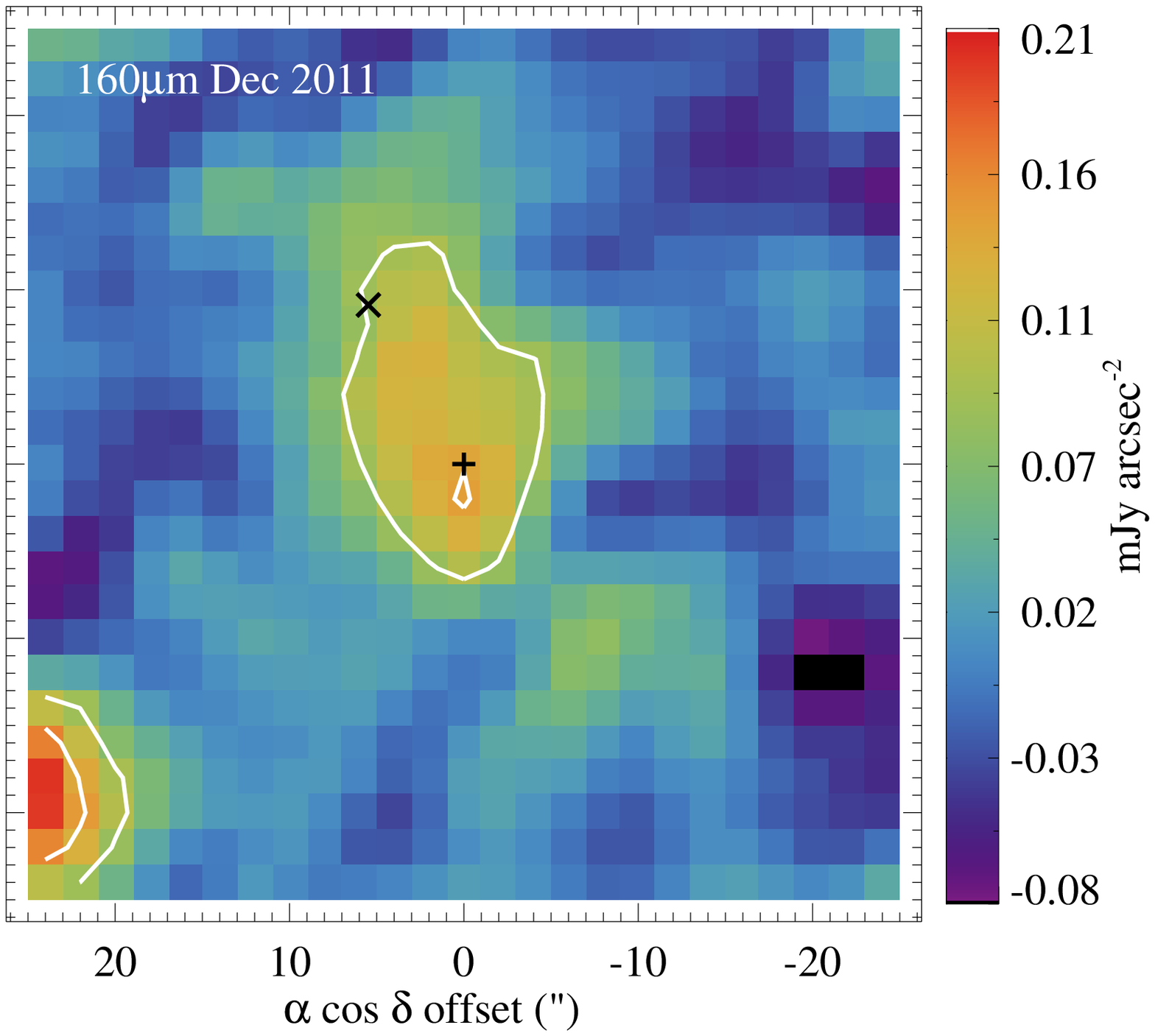} \\
    \vspace{-0.25cm}
    \caption{\emph{Herschel} PACS images of HD~20794, North is up and East is
      left. Contours are at 3, 5, 10, and 25 times the pixel RMS level for each
      observation, though the higher levels are not visible in all images. The surface
      brightness in the scale bars applies to both images in each row. \emph{Top row:} at
      70 $\mu$m, the left panel shows the image and the star has been subtracted from the
      right panel. \emph{Middle row:} at 100 $\mu$m, the left panel shows the image and
      the star has been subtracted from the right panel. \emph{Bottom row:} at 160
      $\mu$m, the left panel shows the image taken with the 100 $\mu$m image, and the
      right panel shows the image taken with the 70 $\mu$m image (neither is
      star-subtracted). The ``+'' symbols shows the star position in each panel, and the
      ``$\times$'' symbols shows the location of a background object. The high Eastward
      proper motion means that the background object is shifted about 2\arcsec~to the
      West (right) in the later 70/160 $\mu$m images relative to the earlier 100/160
      $\mu$m images.}\label{fig:20794ims}
  \end{center}
\end{figure*}

In contrast to HD~38858, most of the emission observed near HD~20794 with PACS comes from
the star itself. The 70, 100, and 160 $\mu$m images of HD~20794 are shown in
Fig. \ref{fig:20794ims}. The top row of panels show the 70 $\mu$m image, where the star
has been subtracted in the right panel. The middle row of panels is the same, but shows
the 100 $\mu$m image. The bottom row shows the two 160 $\mu$m images, the image in the
left panel was taken at the same time at the 100 $\mu$m image, the right panel at the
same time as the 70 $\mu$m image. The star has not been subtracted from these bottom
panels.

It is clear from the 70 and 100 $\mu$m star-subtracted panels in Fig. \ref{fig:20794ims}
(i.e. top right and middle right) that removing the stellar flux density does not account
for all of the emission located at the stellar position. There is also evidence that the
residual emission at the star location is resolved, as the region enclosed by the
3$\sigma$ contour in the star-subtracted 100 $\mu$m image is elliptical with a position
angle pointing slightly East of North. We return to the issue of whether the emission is
resolved when modelling the disc below in section \ref{ss:20794mods}.

The 100 and 160 $\mu$m images also show that there is significant emission that lies to
the NE of the star, as marked by the ``$\times$'' symbol. HD~20794 has a high proper
motion of 3.0\arcsec~yr$^{-1}$ Eastwards and 0.7\arcsec~yr$^{-1}$ Northwards
\citep{2007A&A...474..653V}, so the nearly year long interval between the 100 and 70
$\mu$m observations means that a background object would move Westward (right) relative
to the star by about 2.3\arcsec~between epochs. This movement is not easily discernible
for the NE background object in these images due to a low S/N, but is possibly evident
from both the change in separation and position angle of the stellar and background
source emission in the 160 $\mu$m images.

To measure the flux density of the star+disc emission from HD~20794 we use point-spread
function (PSF) fitting. Though the disc may be marginally resolved, it is also confused
to some degree by the source 10\arcsec~away to the NE. This method gives a measurement
that is more robust to confusion, but that will underestimate the flux density slightly
if the disc is resolved. For the PSF fitting we find $100 \pm 5$ mJy at 70 $\mu$m,
$55 \pm 3$ mJy at 100 $\mu$m, and $27 \pm 3$ mJy at 160 $\mu$m, including calibration
uncertainties. Previous measurements at 70 $\mu$m using MIPS found $94 \pm 14$ and
$107 \pm 4$ mJy \citep{2006ApJ...652.1674B,2013ApJ...768...25G}, consistent with our
measurement. Our values are also consistent with those presented in
\citet{2012MNRAS.424.1206W}. We revisit the derived fluxes when modelling the disc in
section \ref{ss:20794mods}.

\section{Disc models}\label{s:mods}

We now consider the properties of the debris discs around HD~38858 and HD~20794. We have
different levels of information about each, with the HD~38858 disc well resolved and
bright, but the HD~20794 disc faint and only marginally resolved at best, so the
modelling has a level of complexity that reflects this. We first consider the
implications of the flux distributions inferred from the star+disc photometry, and then
move onto modelling of the resolved images. For HD~38858 we also use the disc models to
interpret the upper limit on CO emission and the mass of gas present.

\subsection{HD~38858}\label{ss:38858mods}

\subsubsection{SED}\label{sss:hd38858sed}

\begin{figure}
  \begin{center}
    \hspace{-0.25cm} \includegraphics[width=0.48\textwidth]{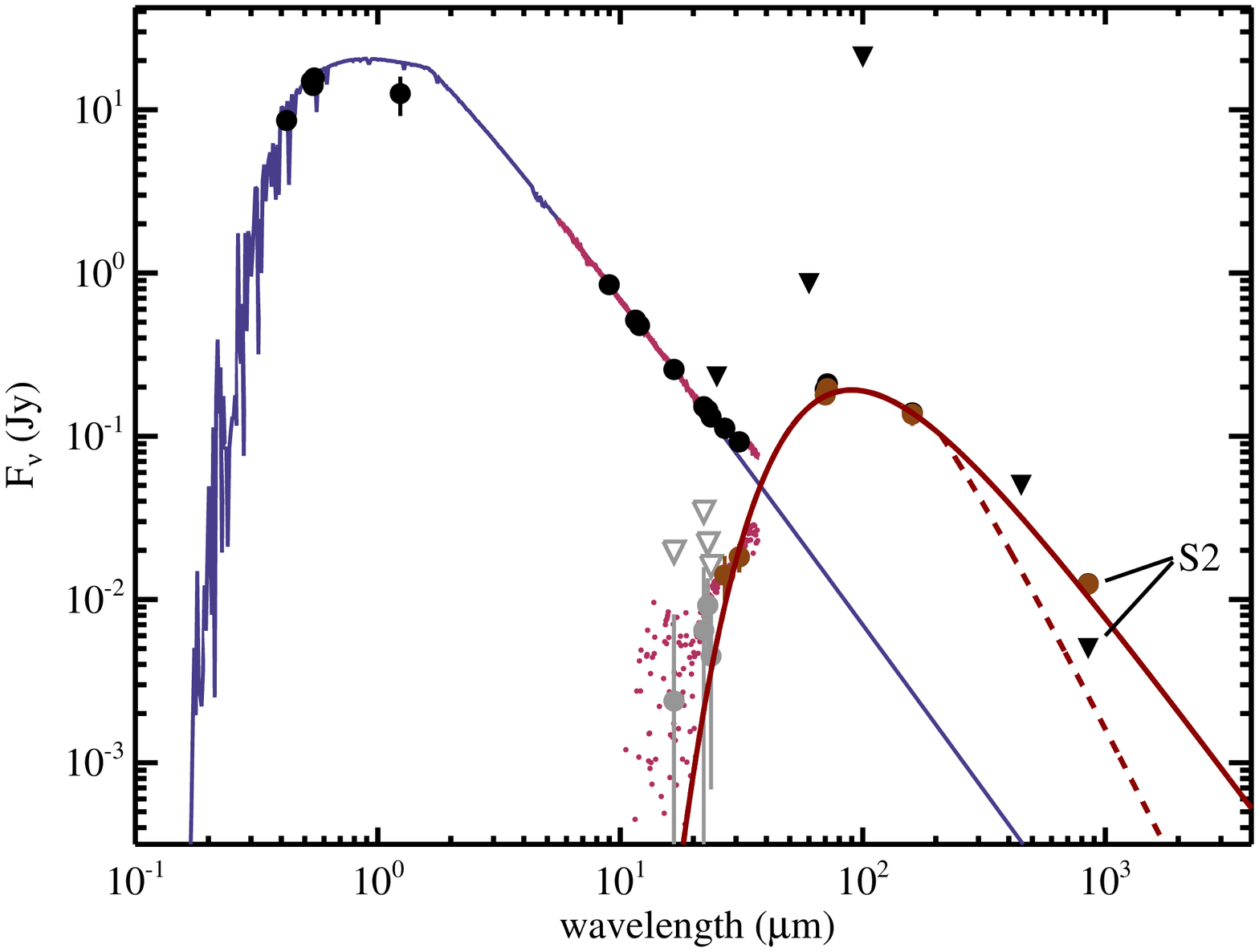}
    \caption{SED for HD~38858. Dots are fluxes and triangles 3$\sigma$ upper
      limits. Black symbols are measured fluxes and brown symbols are star-subtracted
      fluxes (i.e. disc fluxes, which may cover the measured photometry). Grey dots are
      disc fluxes that are consistent with zero, and grey triangles 3$\sigma$ upper
      limits on these fluxes. The 5780K stellar photosphere model is shown in blue, the
      60K blackbody disc model in red. The disc model is a modified blackbody, which
      assumes a SCUBA-2 850 $\mu$m disc flux of $7.5 \pm 2$ mJy.}\label{fig:38858sed}
  \end{center}
\end{figure}

We begin the modelling of HD~38858 with the flux density distribution, which we
abbreviate to ``SED'' for simplicity. The SED from optical to sub-mm wavelengths is shown
in Fig. \ref{fig:38858sed}. The short-wavelength photometry comes from various sources,
and includes ``heritage'' Stromgren and UBV bands
\citep{1998A&AS..129..431H,2006yCat.2168....0M} and space-based Tycho bands
\citep{2000A&A...355L..27H}, but the quality of the photospheric prediction ultimately
relies on mid-IR photometry from IRAS, AKARI, and WISE
\citep{1984ApJ...278L...1N,2010A&A...514A...1I,2010AJ....140.1868W}. HD~38858 was
observed with the IRS spectrograph, and we use the spectrum from the CASSIS database
\citep{2011ApJS..196....8L}. We also include \emph{Spitzer} MIPS photometry from the
aforementioned sources.

To construct an SED model, we fitted photometry shortward of 15 $\mu$m with an AMES-Cond
stellar photosphere model \citep{2005ESASP.576..565B}, finding a stellar effective
temperature of $5780 \pm 20$K and luminosity of 0.83 $L_\odot$. With only the 5-35 $\mu$m
IRS spectrum, and far-IR photometry at 70 and 160 $\mu$m, the spectral coverage for the
disc is poor, which limits our ability to constrain the disc spectrum. For the disc we
therefore simply subtract the expected stellar flux in each band, and fit the remaining
fluxes with a blackbody disc model.\footnote{The AMES-Cond model spectra are computed to
  $\sim$1mm, so the few percent decrease in the stellar 100 $\mu$m flux density relative
  to a Rayleigh-Jeans extrapolation from the mid-IR is included in our photospheric
  prediction.} The temperature of this disc model is well constrained at $60 \pm 2$K.

The long-wavelength part of the disc spectrum depends on our interpretation of the
SCUBA-2 observations (see section \ref{sss:s2}). Assuming that the 450 $\mu$m emission is
not associated with HD~38858, and that the 850 $\mu$m emission is partly contaminated, we
find a disc flux of 7.5 mJy, and the disc spectrum is near to a pure blackbody, as shown
in Fig. \ref{fig:38858sed}. We use a ``modified'' blackbody, where we multiply the
spectrum by $(\lambda_0/\lambda)^\beta$ beyond $\lambda_0$ $\mu$m, with $\lambda_0=210$
$\mu$m and $\beta=0.25$. With only a single measurement beyond 160 $\mu$m, $\lambda_0$
and $\beta$ are degenerate, and their accuracy is also limited by the systematic
uncertainty introduced by our interpretation of the SCUBA-2 850 $\mu$m image.

Useful disc properties that can be derived from the SED are the disc radius assuming
blackbody grains, which yields 21 au (about 3\arcsec), and the disc fractional luminosity
$f = L_{\rm disc}/L_\star$, which is $8.8 \times 10^{-5}$. As can already be seen from
Fig. \ref{fig:38858ims}, the disc is much larger than this estimate, indicating that the
grains in the disc are hotter than blackbodies for a given location, as expected for
grains that emit poorly at long wavelengths. Thus, the requirement of a nearly pure
blackbody spectrum to account for the SCUBA-2 flux is surprising, since if the disc
material lay at a single radius a blackbody spectrum implies that the blackbody and
resolved disc sizes would agree.

However, blackbody disc spectra have been seen for other discs that appear at larger
radii that their blackbody temperature suggest. In the case of AU Mic, the likely
explanation is that the sub-mm observations detect the parent body distribution, which
may have blackbody properties, but that far-IR observations detect a halo of smaller
grains that reside on high eccentricity orbits due to radiation and/or stellar wind
forces \citep[e.g.][Matthews et
al. submitted]{2006ApJ...648..652S,2007ApJ...670..536F}. Therefore, a discrepancy between
the blackbody and observed disc sizes does not warrant the exclusion of sub-mm fluxes
that result in near blackbody disc spectra.

Using the disc temperature and the SCUBA-2 observations, we can also derive the dust mass
in the disc. Assuming an 850 $\mu$m opacity of 45 au$^2$ $M_\oplus^{-1}$ and flux of 7.5
mJy the dust mass is 0.008 $M_\oplus$. Because most bodies are expected to be much larger
than the detected dust, the total disc mass is much larger, depending on the size
$D_{\rm c}$ (in km) of the largest planetesimals. Using equation (15) of
\citet{2008ARA&A..46..339W} and assuming a disc radius of 60 au and a minimum grain size
of 1 $\mu$m yields a total disc mass $M_{\rm tot} \approx D_{\rm c}^{0.5} M_\oplus$, so
about an Earth mass for 1 km sized planetesimals. By assuming a stellar age and that the
debris disc has been evolving for all of this time, $D_{\rm c}$ can be estimated, for
example using equation (16) of \citet{2008ARA&A..46..339W}, where the important
proportionality here is $D_{\rm c} \propto t_{\rm age}^2$. If we assume an age for
HD~38858 of 1 Gyr and the planetesimal properties found for Sun-like stars by
\citet{2011MNRAS.414.2486K}, the maximum size of objects that must be present to
replenish the disc over the star's lifetime is 0.04 km, though this size is very
uncertain because it depends on the square of the stellar age and planetesimal
strength. For weaker planetesimals as found for A-type stars by
\citet{2007ApJ...663..365W} the size is of order 10 km. Thus, the corresponding disc mass
is between about 0.1 to 3 Earth masses, and only depends linearly on the age and object
strength. We can also estimate the rate at which mass is being lost from the disc as the
total mass divided by the age, which yields $10^{15}$ to $10^{16}$ kg yr$^{-1}$. Because
the derived mass depends on the age (via $D_{\rm c}$), the mass loss rate is independent
of the assumed age and the main contributor to the uncertainty as calculated here is the
planetesimal strength. This mass loss rate will be used when considering the mass of CO
gas below in section \ref{ss:gas}.

% print,(1e3/(1.4e-9*60^(13/3.)*0.5*150^(5/6.)*0.05^(-5/3.)))^2
% print,(1e3/(1.4e-9*60^(13/3.)*0.5*3700^(5/6.)*0.05^(-5/3.)))^2

\subsubsection{Resolved models}\label{sss:hd38858res}

We now construct resolved models of the PACS images of HD~38858 to constrain the disc
spatial structure. For a given disc structure, we first generate high resolution images
at each observed wavelength, and then convolve these with the telescope response to a
point source (i.e. the PSF, which is a calibration observation of $\gamma$ Dra) for
comparison with the observations. All of our Herschel images are created with North
pointing upwards, so the PSF observations, which have a different telescope orientation,
must be rotated to the correct orientation.

\begin{figure*}
  \begin{center}
    \hspace{-0.25cm} \includegraphics[width=1\textwidth]{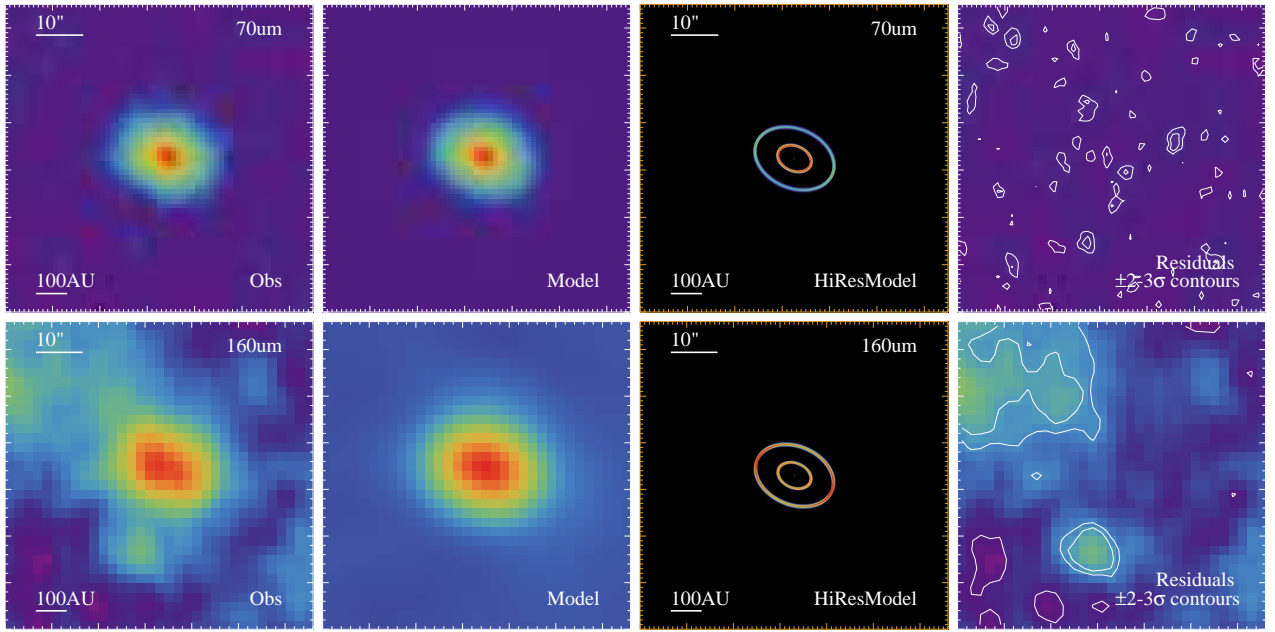}
    \caption{Two ring model of the HD~38858 disc. The columns show the PACS image,
      convolved model, high resolution model, and residuals from left to right. The top
      row is at 70 $\mu$m, and the bottom row 160 $\mu$m. The model has two discrete
      rings that are both 10 au wide, and are centred at 55 and 130 au. This model fits
      the PACS data very well, with the only significant residuals being due to the high
      background level at 160 $\mu$m.}\label{fig:38858rings}
  \end{center}
\end{figure*}

The disc is well resolved and our unsuccessful attempts to model the 70 $\mu$m image with
a single narrow ring find that the disc has significant radial extension. We therefore
use two similar models, one in which the disc is simply made up of two narrow dust belts,
and one where the disc is radially extended. Common parameters for both models are the
disc inclination $i$ and position angle $\Omega$ (East of North), and the disc
temperature profile as defined by $T_{\rm disc}=f_{\rm T} T_{\rm BB}$. The factor
$f_{\rm T}$ accounts for how much hotter the dust appears compared to a blackbody at the
same distance from the star, where $T_{\rm BB}=278.3 L_\star^{1/4}/\sqrt{r}$. That is,
$f_T$ reconciles the disc spatial structure with the SED. All models are consistent with
$f_{\rm T}=1.9$, similar to the values found for the disc around the Sun-like star 61 Vir
\citep{2012MNRAS.424.1206W}. We use modified blackbody parameters for the disc spectrum
as described above, though here our goal is to model the disc structure as seen by PACS
so they are relatively unimportant.

\begin{figure*}
  \begin{center}
    \hspace{-0.3cm} \includegraphics[width=0.5\textwidth]{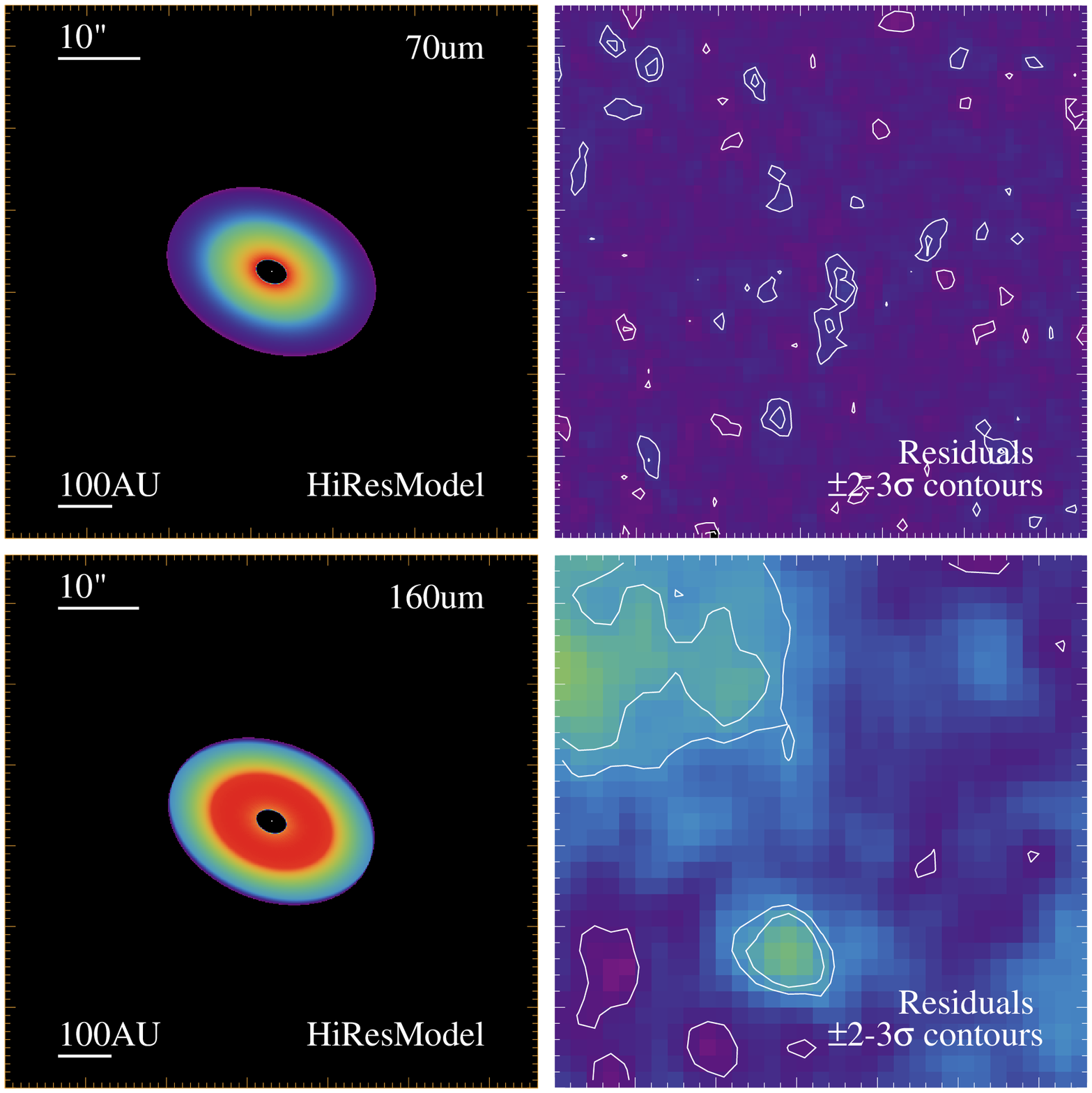}
    \hspace{-0.1cm} \includegraphics[width=0.5\textwidth]{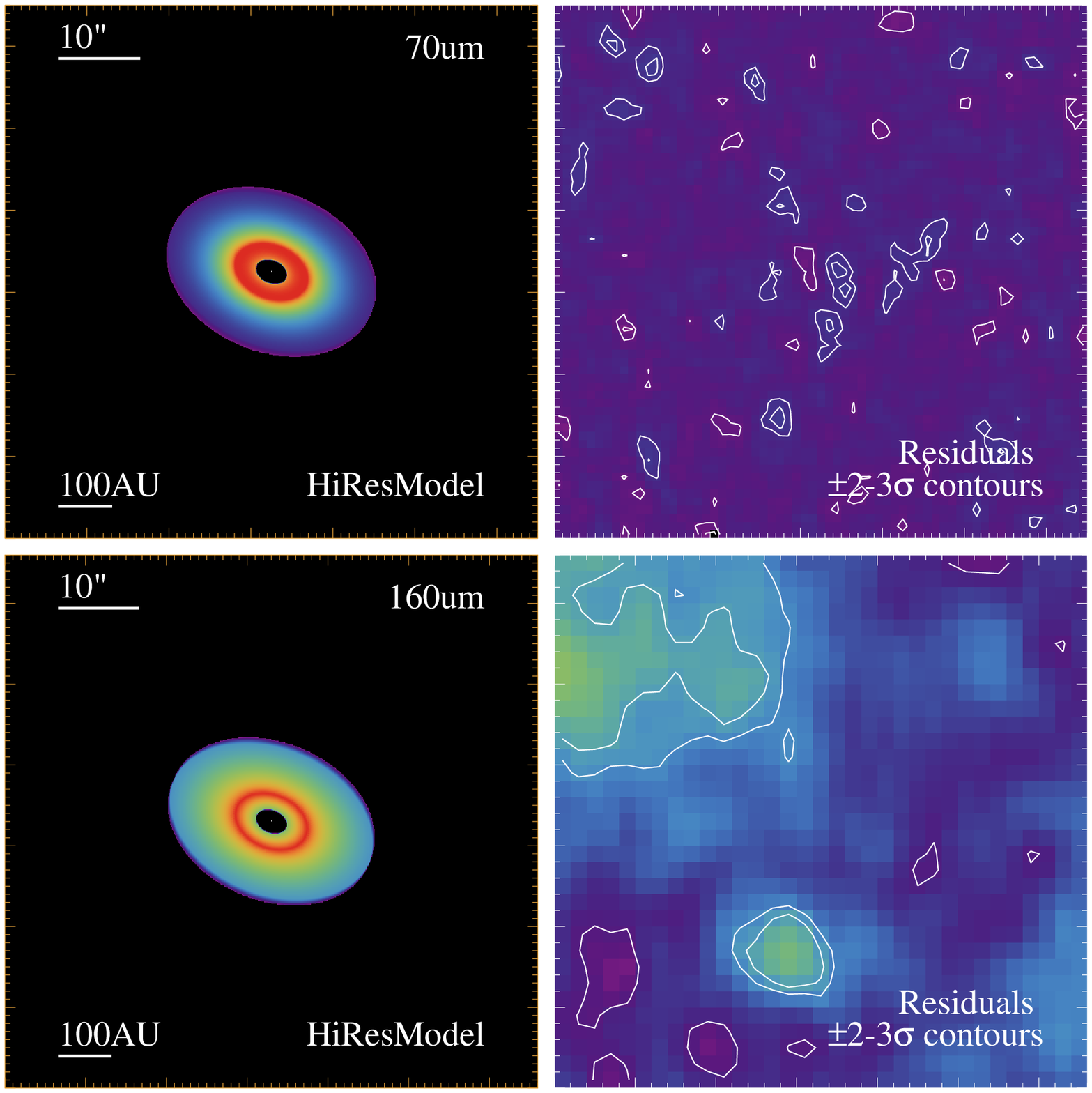}
    \caption{Two alternative models of the HD~38858 disc. Each square of four panels
      shows the high resolution model and residuals at 70 and 160 $\mu$m. The left set of
      panels shows a model where the optical depth profile increases from 30 to 117 au,
      and then decreases out to 200 au. The right set of panels shows a model where the
      optical depth increases from 30 to 70 au, as is then flat out to 200
      au.}\label{fig:38858cont}
  \end{center}
\end{figure*}

For the two ring model we use belts that are 10 au wide with constant optical depths, and
the best fitting model is shown in Fig. \ref{fig:38858rings}. The disc inclination and
position angle are 44$^\circ$ and 67$^\circ$, with estimated uncertainties of
5$^\circ$. These measurements are consistent with the geometry reported by
\citet{2012AJ....144...45K} from lower resolution \emph{Spitzer} observations. The two
rings are centred on 55 and 130 au, with face-on optical depths of $4.5 \times 10^{-4}$
and $1.3 \times 10^{-3}$. The observed-model residuals in the right panels of
Fig. \ref{fig:38858rings} show that this model is a good fit to the PACS observations,
with the only significant remaining structure attributable to the high background
level. The model also reproduces the disc spectrum well, though a lack of far-IR
photometry means that this goal is easily met.

We also considered two similar extended models for the disc. These consist of two disc
zones, that extend from $r_{\rm in}$ to $r_{\rm mid}$, and then from $r_{\rm mid}$ to
$r_{\rm out}$. These two zones have independent power-law optical depth profiles $\tau
\propto r^{\alpha_{\rm i}}$, where $i$ is 1 or 2 for the inner and outer zones. The
absolute level of the zones is set such that the join at $r_{\rm mid}$ is smooth. The
disc inclination and position angles are as in the two ring model, and allowing them to
vary does not improve the model fit.

For a first model, we set $\alpha_1=1$ and $\alpha_2=-1$, so the optical depth increases
from some inner edge to a peak value, and then decreases to the outer edge. For these
assumptions we find $r_{\rm in}=30$, $r_{\rm mid}=117$, and $r_{\rm out}=200$ au. To
explore the parameter uncertainties we construct a second model with $\alpha_1=1.7$ and
$\alpha_2=0$, for which we find $r_{\rm in}=30$, $r_{\rm mid}=70$, and $r_{\rm out}=200$
au. These two models are shown in Fig. \ref{fig:38858cont}, which both have similarly
good residual images, indicating that neither of these models should be preferred
relative to each other, or relative to the two ring model described above. This two zone
model clearly has some degree of degeneracy, with $r_{\rm mid}$ shifting to account for
changes in the optical depth profiles. The inner and outer radii are common however,
suggesting that for continuous models the inner edge is around 30 au, and the outer edge
around 200 au.

In comparison to the two ring model, the two zone model extends to both smaller and
larger radii. This is to be expected however; a model that was continuous only between
the two radii found by the two ring model would have too much emission from the region
between the ring locations, so this concentration is countered by making the continuous
model cover a wider range of radii.

We briefly compare the prediction of our model for the scattered light brightness with
limits set by \citet{2012AJ....144...45K} using the Space Telescope Imaging Spectrograph
(STIS) instrument onboard the Hubble Space Telescope (HST). The limits decrease with
increasing radius, being about 21 mag arcsec$^{-2}$ at 3.5\arcsec~(55 au) and 23.3 mag
arcsec$^{-2}$ at 8.5 \arcsec~(130 au). If we assume isotropic scattering and an albedo of
0.1, the inner component of our two belt model is too faint to detect (22 mag
arcsec$^{-2}$), and the outer component is near the limit of detectability. Therefore the
upper limit on the albedo is about 0.5 for the inner belt, and about 0.1 for the outer
belt. For the continuous model the surface brightnesses are about one magnitude fainter
for an albedo of 0.1, and the inner regions of such a disc cannot be detected even with
an albedo of 0.9, at which point the outer disc is at the limits of detectability. Low
albedoes are generally the norm for scattered light observations of debris discs
\citep[e.g.][]{2010AJ....140.1051K}, with relatively strong forward scattering being a
possible explanation \citep{2010A&A...509L...6M,2013A&A...549A.112M}. Therefore, the STIS
observations do not set sufficiently strong limits that the dust albedo in the HD~38858
disc appears any different to discs that were detected in scattered light.

In summary, we find that several different models can reproduce the structure in the
HD~38858 disc. It is not clear from the PACS data whether the disc is split into two
separate components at about 55 and 130 au, or whether the disc covers a wide region from
about 30 to 200 au. For the latter models, the optical depth profile is not clear because
it is possible to construct a two zone model with differing profiles in each zone,
depending on the radial location of the join between these zones.

\subsubsection{SCUBA-2 interpretation}\label{sss:s2mod}

\begin{figure}
  \begin{center}
    \hspace{-0.025cm} \includegraphics[width=0.45\textwidth]{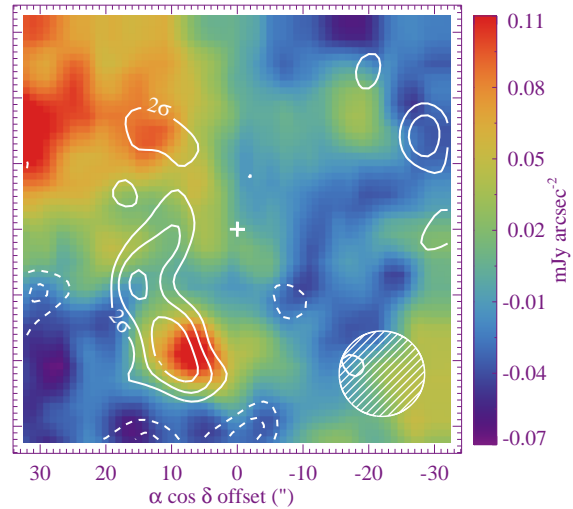}
    \caption{Residual 850 $\mu$m SCUBA-2 image (contours), with residual 160 $\mu$m image
      (colours) in the background. The contours are at 2, 3 and 4$\sigma$ (solid are +ve,
      dashed are -ve). The star location is indicated by the white cross. The SCUBA-2 850
      $\mu$m beam (13\arcsec) is shown at the lower right.}\label{fig:s2mod}
  \end{center}
\end{figure}

As noted in section \ref{sss:s2}, our SCUBA-2 images appear to suffer from some
background contamination, most likely from the Orion Complex. As suggested by the
contours in Fig. \ref{fig:38858s2}, the 850 $\mu$m image likely also contains flux
originating in HD~38858's debris disc. Having derived possible disc structures using the
PACS data, we can therefore assume that the structure is similar at 850 $\mu$m and
attempt to verify this interpretation.

Using the models of the disc structure derived above (which all yield similar results),
Fig. \ref{fig:s2mod} shows 850 $\mu$m residual contours after subtracting a model with a
total disc flux of 7.5 mJy. The model was created in the same way as the PACS models, but
at a wavelength of 850 $\mu$m and using a 13\arcsec~Gaussian for the PSF. The residuals
show a ``stripe'' of emission to the East of the stellar position, which extends between
two regions where significant residual emission remains in the 160 $\mu$m image (see also
Figs. \ref{fig:38858rings} and \ref{fig:38858cont}). The choice of total disc flux is
somewhat subjective, and was chosen so that the stripe is fairly uniform in brightness
from North to South; we assign an uncertainty of 2 mJy (i.e. 3$\times$ higher than the
point source uncertainty). That we can successfully subtract a symmetric disc model
centred on the expected stellar position, and that the residual flux has good
correspondence with the 160 $\mu$m residual emission, shows that our interpretation is
reasonable. It may of course be that the disc is asymmetric and some of the residual 850
$\mu$m flux to the East actually belongs to the disc. However, given that the PACS
images, that have much higher S/N, are consistent with being symmetric, we consider this
possibility unlikely.

\subsubsection{Gas mass}\label{ss:gas}

To estimate the upper limit on the total CO gas mass from our limit on the integrated
J=2-1 flux (section \ref{sss:co}), we apply the CO excitation model described in
\citet{2015MNRAS.447.3936M}. The model solves the statistical equilibrium of CO
rotational levels, taking into account both collisional and radiative excitation in a
full non-local thermodynamic equilibrium (NLTE) approach. The radiation field impinging
on CO molecules at the frequency of the transition J=2-1 (230.538 GHz) is dominated by
the cosmic microwave background (CMB), with negligible contribution from both the star
(beyond about 1 au) and from any of the disc models described above (regardless of
location). That is, the radiative excitation of CO molecules is largely independent of
their radial location. Additional excitation of the molecules by other means, most likely
collisions with electrons freed from carbon ionisation following CO dissociation, will
only lead to more J=2-1 emission from a fixed mass of CO. We can therefore set a hard
upper limit on the total CO mass around HD~38858 of 1.5 $\times$ 10$^{-4}$ M$_{\oplus}$.

% print,bnuw(1300,2.7)
% CMB 3.17e-18 W/m^2/hz/sr = 3.17e8 Jy/sr = 4e9 Jy from whole sky
% star = 3.82e-5 at 1300um at 15.2pc -> find r where the same
% print,!pi*bnuw(1300,5800)*asin(3.88e-3/1.2)^2

The age of HD~38858 is well beyond when gas of any kind has been detected around young
stars \citep{1995Natur.373..494Z,2011ApJ...740L...7M,2015MNRAS.447.3936M}, so little is
expected to exist in this system, and any gas that is present likely originates in icy
planetesimals within the observed debris disc \citep[e.g.][]{2012ApJ...758...77Z}. If we
assume such a second-generation scenario where the gas is composed of cometary molecules
such as H$_2$O, CO and their photodissociation products, electrons (produced via CO
photodissociation followed by C ionisation) are the main colliders and we can set
stricter limits on the total CO mass. Such a constraint can be illustrated by assuming a
value of 100 for the C/CO ratio \citep[as observed in $\beta$
Pictoris,][]{2000ApJ...538..904R}, and an ionisation fraction C$^+$/C of 1.7 \citep[as
estimated by][at 100 au around a G-type star]{2006ApJ...643..509F}. For HD38858, the
interstellar UV field dominates photodissociation of CO even at the inferred inner edge
of the dust disc (30 au), yielding a CO lifetime of $\sim$120 years, while H$_2$O will be
affected by the stellar radiation field, dissociating in 3 years at 30 au, and in 40
years at 200 au. For gas released from planetesimals with Solar System cometary CO/H$_2$O
abundance ratios \citep[between 0.4 and 30\%,][]{2011ARA&A..49..471M}, electrons dominate
the collisional excitation of CO. We then find a CO mass upper limit of 3.2 $\times$
10$^{-5}$ M$_{\oplus}$. While this limit depends on the above assumptions, it shows how
the limit on gas can change with the scenario assumed.

% Below CO/H$_2$O = X, SOMETHING starts to become important for the excitation of CO, in
% which case the mass limits become lower.

% This leads to a lower CO mass upper limit of 3.8 $\times$ 10$^{-5}$ M$_{\oplus}$, where
% this is independent of our assumptions about collisional partners, since adding more, or
% using different gas temperatures or different CO/H$_2$O abundance ratios would only push
% the gas towards LTE and hence yield lower CO mass limits.

We then consider what this upper limit means in terms of production and destruction of
secondary gas in the system. CO is mainly destroyed via photodissociation by the
interstellar UV, with a timescale of $\sim$120
years. %Assuming the disc is optically thin to UV radiation, the CO photodissociation timescale is dominated by the interstellar radiation field (ISRF) at all disc radii, and is estimated to be \citep[$\sim$120 years][]{Visser2009}.
Dividing the upper limit on CO mass by this timescale, we set an upper limit on the CO
production rate of 1.6 $\times$ 10$^{18}$ kg
yr$^{-1}$. %These are shown as solid coloured lines in Fig. X.
On the other hand, we can estimate the CO production rate from the mass loss rate of
solids through collisions in the disc (10$^{15}$ to 10$^{16}$ kg yr$^{-1}$, section
\ref{sss:hd38858sed}), where the production rate depends on the ice/rock and CO/H$_2$O
fractions. For Solar-System-type comets, we therefore obtain an observational upper limit
on the CO production rate from the JCMT observations that is at least two orders of
magnitude higher than the upper limit from the derived planetesimal destruction
rate. Thus, the limit on CO gas is not stringent enough to test different scenarios for
the production of gas by planetesimal collisions.

\subsection{HD~20794}\label{ss:20794mods}

For HD~20794 we first consider the SED, and then show what the non-detection of excess at
mid-IR wavelengths implies for the radial location of the dust detected by near-IR
interferometry. We then turn to the images and consider the disc spatial structure.

\subsubsection{SED}

\begin{figure}
  \begin{center}
    \hspace{-0.25cm} \includegraphics[width=0.48\textwidth]{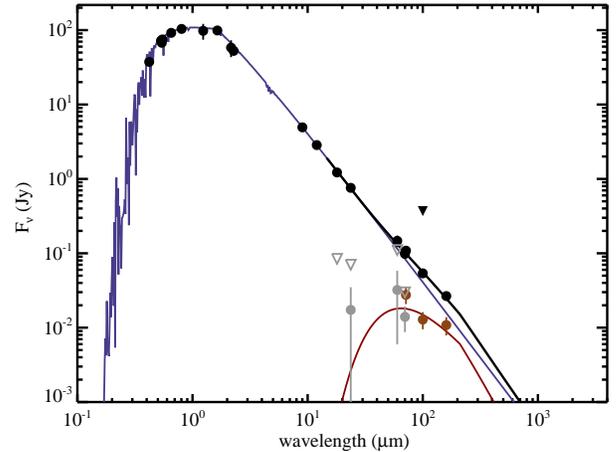}
    \caption{SED for HD~20794. Dots are fluxes and triangles 3$\sigma$ upper
      limits. Black symbols are measured fluxes and brown symbols are star-subtracted
      (i.e. disc) fluxes. Grey dots are disc fluxes that are consistent with zero, and
      grey triangles 3$\sigma$ upper limits on these fluxes. The 5480K stellar
      photosphere model is shown in blue, the 80K blackbody disc model in red, and the
      star+disc spectrum in black.}\label{fig:20794sed}
  \end{center}
\end{figure}

We follow the same procedure described above to model the HD~20794 spectrum, shown in
Fig. \ref{fig:20794sed}. The photospheric prediction in the far-IR is well constrained by
the IRAS and AKARI observations. We find a stellar effective temperature of $5480 \pm
20$K, and a disc temperature of $80^{+70}_{-30}$K. The disc temperature is very poorly
constrained due to the lack of disc detections over a range of wavelengths and because
the disc is fairly faint. Despite the large temperature uncertainty, the disc fractional
luminosity is reasonably well constrained to between $10^{-5}$ and $10^{-6}$, because
disc models that fit the disc spectrum have similar integrated luminosities. The lower
end of this range is only an order of magnitude above the Solar System level of $10^{-7}$
\citep{2012A&A...540A..30V}.

Based on the disc temperature, the blackbody disc radius is about 10 au, or 3\arcsec~in
diameter, though the uncertainty is very large because the blackbody radius is
proportional to the inverse square of the temperature. This size prediction suggests
however that the disc may be resolved, as was also suggested by the images in
Fig. \ref{fig:20794ims}. We now consider the location of dust detected with near-IR
interferometry, and then turn to a more detailed consideration of the PACS images to test
whether the disc is indeed resolved, and if so, whether its properties can be constrained
further.

\subsubsection{Location of the hot dust}\label{sss:20794hot}

A near-IR excess at 1.65 $\mu$m ($H$-band), with a disc/star flux ratio of
$1.64 \pm 0.37$ percent, was recently detected around HD~20794 by
\citet{2014A&A...570A.128E}. Such excesses are fairly common and are thought to arise
from hot dust located very near to the parent star
\citep[e.g.][]{2011A&A...534A...5D,2013A&A...555A.104A,2013A&A...555A.146L}. Their origin
remains a mystery; dust should be removed by radiation forces on orbital timescales, and
local planetesimal populations that resupply the dust would be rapidly depleted to
undetectable levels by collisions. There is also no support for a non-local origin, in
particular because the hot dust brightness does not correlate with other system
properties such as the presence of outer dust belts \citep[but may correlate positively
with stellar age,][]{2013A&A...555A.104A,2014A&A...570A.128E}. It may be that the
phenomenon is more closely related to stellar astrophysics, for example being very small
dust trapped by the stellar magnetic field \citep{2013ApJ...763..118S}, though in such a
scenario the dust may still be supplied by known debris disc processes.

Among the systems detected by \citet{2014A&A...570A.128E}, the spectral slopes seen
across the $H$-band (1.55 to 1.75 $\mu$m) are relatively flat compared to the stellar
flux. In the case of HD~20794 dust cooler than about 500 K could be ruled out, and while
hot dust remains a possibility, the spectrum is also consistent with arising from
scattered starlight. This latter possibility means that the dust could lie anywhere
within the 400 mas PIONIER field of view, and for a given surface area of dust therefore
extend farther from the star than if the emission is thermal, perhaps even into the
habitable zone. No excess was seen at 24$\mu$m for HD~20794, so the upper limit on the
dust seen at this wavelength provides useful constraints on where the hot dust
lies. Deriving this constraint however requires exploration of the grain properties, in
particular the albedo and the emission or scattering efficiencies at $H$-band relative to
the emission efficiency at 24 $\mu$m.

\begin{figure}
  \begin{center}
    \hspace{-0.25cm} \includegraphics[width=0.48\textwidth]{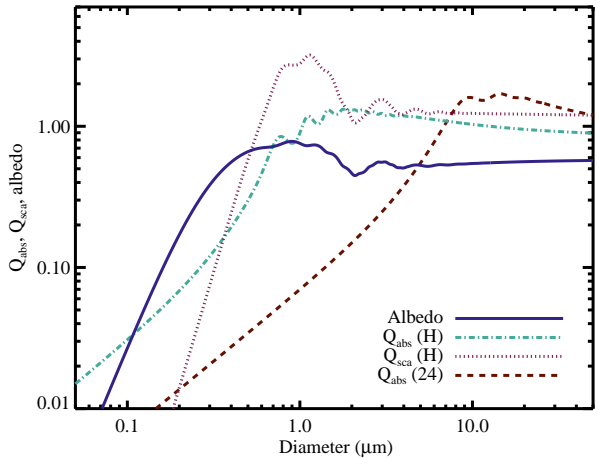}
    \caption{Scattering and absorption properties for silicate/organic grains as a
      function of grain diameter. The dotted and dot-dash lines show $Q_{\rm abs}$ and
      $Q_{\rm sca}$ at H-band, and the dashed line shows $Q_{\rm abs}$ at 24 $\mu$m. The
      solid line shows the albedo $Q_{\rm sca}/(Q_{\rm sca}+Q_{\rm abs})$. The H-band
      albedo relative to the 24 $\mu$m emission efficiency is maximised for grains about
      0.5 $\mu$m in size.}\label{fig:albedo}
  \end{center}
\end{figure}

\begin{figure}
  \begin{center}
    \hspace{-0.25cm} \includegraphics[width=0.48\textwidth]{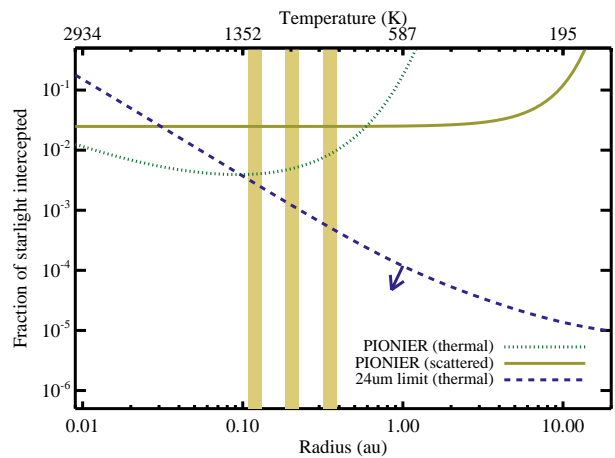}
    \caption{Limits on the fraction of starlight intercepted and the location of 0.5
      $\mu$m size dust grains around HD~20794 detected with PIONIER. The near horizontal
      solid line shows a locus of possible dust locations if the detection originates in
      scattered light. The dotted line shows a locus of dust locations if the detection
      originates in thermal emission from blackbodies.  The dashed line shows the
      3$\sigma$ upper limit set by the MIPS 24 $\mu$m non-detection. The vertical stripes
      show the locations of the planets.}\label{fig:20794hot}
  \end{center}
\end{figure}

The grain scattering ($Q_{\rm sca}$) and emission/absorption ($Q_{\rm abs}$) efficiencies
are shown in Fig. \ref{fig:albedo} (the emission and absorption efficiencies are
equal). Lines show these values as a function of grain diameter for non-porous grains
composed of 1/3 silicates and 2/3 organics \citep{1999A&A...348..557A} at H-band and 24
$\mu$m. The albedo, $\omega=Q_{\rm sca}/(Q_{\rm sca}+Q_{\rm abs})$ at H-band, is also
shown. For small grains the albedo becomes very small because grains absorb better than
they scatter, and the maximum ratio between the H-band albedo and the 24 $\mu$m emission
efficiency is reached for grains about 0.5 $\mu$m in size. These properties set limits on
the grain size if the PIONIER emission originates in scattered light, because no more
than 100\% of the starlight can be intercepted by the dust. That is, assuming isotropic
scattering the fraction of starlight intercepted by the dust
$f_{\rm cap}=0.0164/\omega<1$ so from Fig. \ref{fig:albedo} the grains must be larger
than about 0.1 $\mu$m for the assumed composition. As an example the locus of possible
dust locations for grains 0.5 $\mu$m in size is shown as the solid line in
Fig. \ref{fig:20794hot}. Without further constraints, the dust could lie anywhere along
this line, the radius only being restricted by the PIONIER field of view. At each point
along this locus however, a corresponding amount of thermal emission is expected, so some
locations can be ruled out by an analogous locus derived from the 24 $\mu$m
non-detection.

A constraint on the hot dust location based on the 24 $\mu$m non-detection is shown as
the dashed line in Fig. \ref{fig:20794hot}. This line was computed using equation (11)
from \citet{2008ARA&A..46..339W}, but the dust temperature and $Q_{\rm abs}$
($=1/X_\lambda$) were calculated self-consistently for 0.5 $\mu$m grains. The dust must
lie below the dashed line, so if the PIONIER emission originates in light scattered off
0.5 $\mu$m dust with our assumed composition it lies within 0.03 au. For smaller dust the
albedo becomes smaller faster than the decrease in 24 $\mu$m emission efficiency, so the
radial constraint becomes stronger. For larger dust the 24 $\mu$m emission efficiency
increases faster than the albedo so again the radial constraint is stronger.

An additional issue for a scattered light origin is forward (or backward) scattering,
which could be relevant here as we infer particle sizes similar to or larger than the
PIONIER-observed wavelength of 1.65 $\mu$m. For a near edge-on disc the total flux can
exceed that from isotropically scattering dust, and thus the solid line may be
overestimated in Fig. 10. The enhancement in the disc flux required to alter our
conclusions is large however; the solid line would need to move down by two orders of
magnitude for the dust location to be allowed near 1 au. To make a simple estimate, we
computed the integrated emission from a circular dust ring using the
\citet{1941ApJ....93...70H} phase function. For the emission to exceed that compared to
isotropic scattering by an order of magnitude requires very strong forward scattering
($g>0.93$) from a near-perfectly edge-on disc. In this case however, almost all of the
emission originates from a ten-degree arc in the ring. This kind of emission would mimic
the off-centre point-like emission from a companion, rather than that of a
centro-symmetric disc, and the former has already been ruled out for HD~20794
\citep{2014A&A...570A.127M}. This illustration is of course very basic, but shows that
forward scattering is very unlikely to result in a brightness enhancement that is both
consistent with the PIONIER observations, and results in a much smaller fraction of
starlight intercepted by the dust compared to what would be inferred assuming isotropic
scattering. We therefore conclude that our constraints on the dust location (if it
originates in scattered light) are robust.

The green dotted line in Fig. \ref{fig:20794hot} shows the location of the hot dust if it
originates in thermal emission, computed in the same way as at 24 $\mu$m. Inside about
0.5 au the line lies below that for scattered light, so the same signature can be created
from thermal emission but with several times less surface area in dust. The 24 $\mu$m
constraint still applies, so for thermal emission from 0.5 $\mu$m grains the dust lies
within 0.1 au. This constraint is similar for smaller grains; Fig. \ref{fig:albedo} shows
that the ratio of $Q_{\rm abs}$ at H-band and at 24 $\mu$m is similar for grains smaller
than 1 $\mu$m, and thus for smaller dust both lines move upwards in concert. For larger
grains the lines move downwards slightly and the radial constraint moves inwards until
$Q_{\rm abs}=1$, and the constraints then do not change with increasing grain size
(i.e. they behave as black bodies above about 10 $\mu$m in size). These radial
constraints on the dust location are similar for other compositions, for example the
possibility that the grains contain carbon is motivated by the higher sublimation
temperature \citep{2013A&A...555A.146L}. The constraint moves inwards somewhat for
compositions where $Q_{\rm abs}$ at 24 $\mu$m relative to H-band becomes larger than the
example here (e.g. due to a strong spectral feature for small silicate grains).

If the dust is long-lived, which is suggested by the relatively common detection of such
dust around other stars, it would be unlikely to reside near the planets, which would
rapidly accrete or scatter the dust (and any parent planetesimals that could act as a
local mass reservoir). The 24 $\mu$m limits rule out significant H-band emission beyond
the planets, so the dust most likely lies interior to them. Thus, our analysis based on
non-detection of excess emission with MIPS at 24 $\mu$m suggests that the hot dust seen
around HD~20794 with PIONIER lies within $\sim$0.1 au of the star ($\lesssim$20 Solar
radii).

\subsubsection{Resolved models}\label{sss:20794res}

\begin{figure*}
  \begin{center}
    \hspace{-0.25cm} \includegraphics[width=0.54\textwidth]{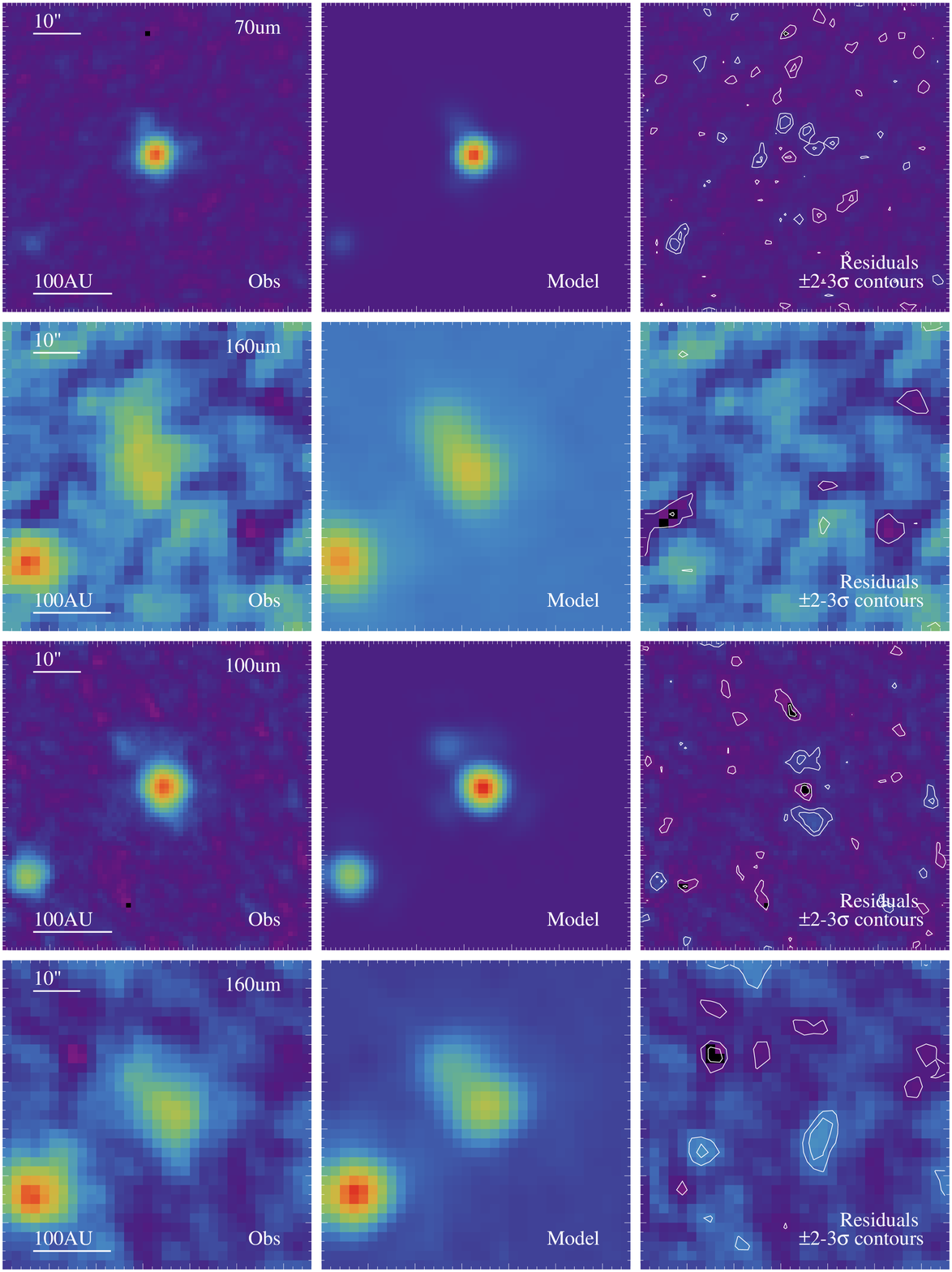}
    \hspace{0.3cm} \includegraphics[width=0.359\textwidth]{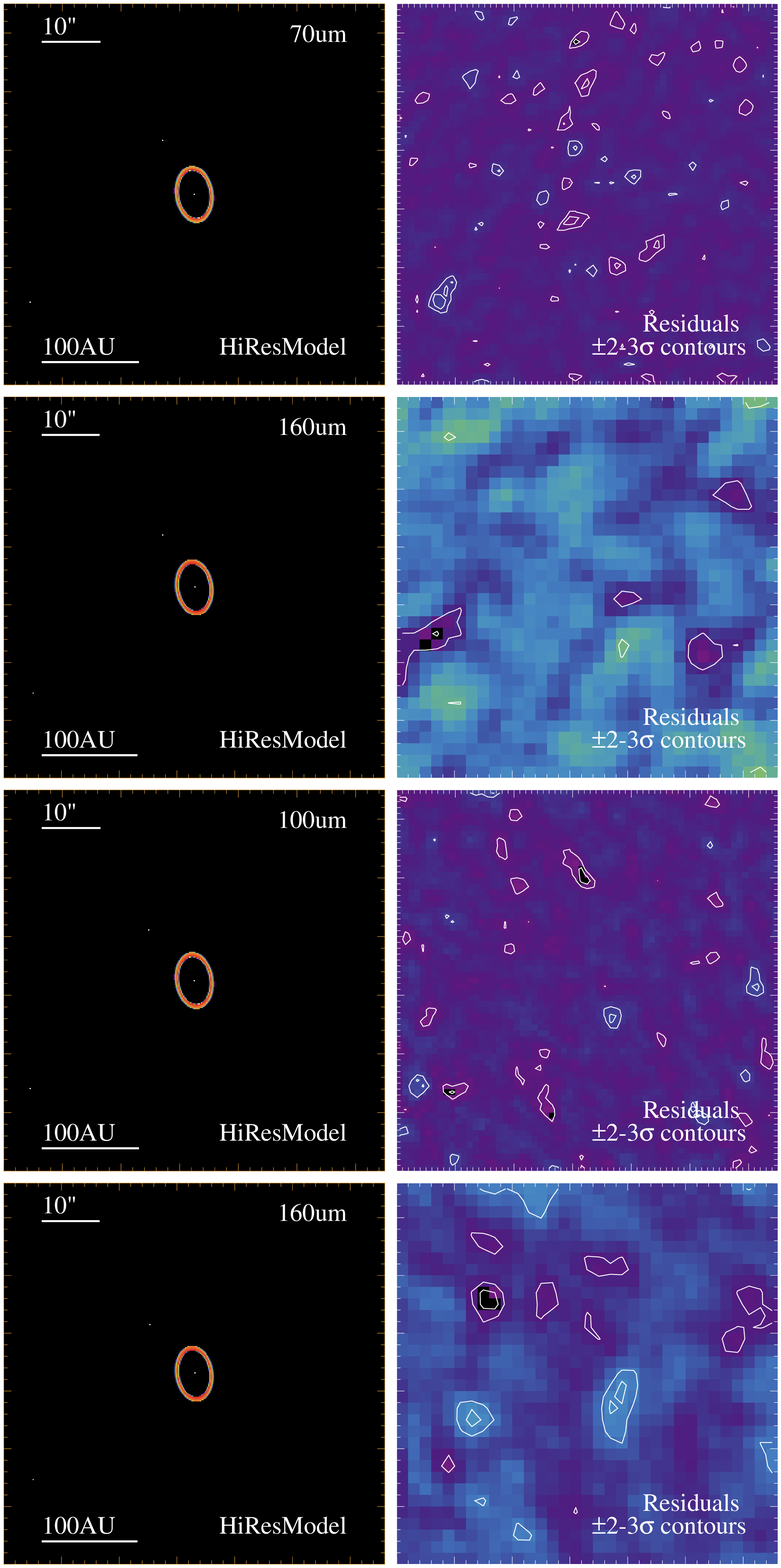}
    \caption{Disc models for HD~20794. Each row of images is a different wavelength, the
      top two rows are for the 70 and 160 $\mu$m observation, and the bottom two rows are
      for the 100 and 160 $\mu$m observation. The left (first) column shows the PACS
      images, the second column shows the model where the disc is assumed to be
      unresolved and the third column shows the residuals for this model (i.e. first
      column - second). The fourth column shows a resolved disc model and the fifth
      column shows the residuals for this model. The point source model leaves residuals
      on either side of the star at 100 $\mu$m, and these are accounted for with the
      resolved disc model.}\label{fig:20794mods}
  \end{center}
\end{figure*}

Returning to the PACS images of HD~20794, we first test whether the disc is resolved by
constructing a point source model of disc emission. The components of this model are a
source at the star location, the NE background source, and the bright SE background
source. As in Fig. \ref{fig:20794ims}, we take the motion of the star between the 100 and
70 $\mu$m epochs into account. The point source at the star location represents the star
plus an unresolved disc. This is in fact the PSF fitting method used to derive the point
source fluxes quoted in section \ref{ss:20794obs}, and the left three columns in
Fig. \ref{fig:20794mods} show the results of this model.
 
The left column shows the observations, the second column the point source model
(convolved with the PACS beam), and the third column the residuals after subtracting this
model from the observations. The best-fit point source fluxes (after subtraction of the
stellar flux) are 10.9, 12.3, and 8.9 mJy at 70, 100 and 160 $\mu$m, where the
uncertainties are the same as the point source uncertainties of 5, 3, and 3 mJy given in
section \ref{ss:20794obs}. In general the model is very good, finding significant fluxes
(i.e. an excess), but notably leaves residuals on either side of the star at 100 $\mu$m,
which is the deepest image.

We therefore allow the disc component to be resolved, modelling it as a simple narrow
ring of width 5 au with free parameters of disc radius, inclination, and position
angle. The best fitting resolved disc model is shown in the two rightmost columns of
Fig. \ref{fig:20794mods}, where the fourth column shows the disc model at high
resolution, and the fifth column shows the residuals for this model. The model has a
radius of 24 au, an inclination of 50$^\circ$, and a position angle of 8$^\circ$ E of N.

This resolved model accounts for the residual flux that was present on either side of the
star for the point source model, but given that this difference lies in only a few pixels
that were about 3$\sigma$ outliers, the extra complication added by a resolved model is
only marginally warranted, and therefore uncertainties on the disc parameters are
moot. Despite this issue, it seems likely that the disc is resolved with a high
inclination, with a diameter of around 50 au (10\arcsec), and as one of the nearest
Sun-like stars with a disc and planets is a high priority for future observations.

Given that both 61~Vir and HD~38858 host discs with significant radial extent, we can
consider what the observations rule out for such extended discs around HD~20794. By
analogy with the two-belt model for HD~38858, we added a second 5 au wide belt at a
radius of 50 au to the resolved HD~20794 model. To remain undetected the optical depth in
this belt must be less than about half the value in the inner belt. For HD~38858 the
outer belt has three times greater optical depth than the inner, and for 61~Vir the two
belts have similar optical depths. A basic conclusion is therefore that the material in
the disc around HD~20794 is the most centrally concentrated of the three.

\section{Planetary system structure}\label{s:disc}

We now discuss our results within the context of the wider planetary system structure. We
begin by presenting a brief update on the radial velocity (RV)-discovered planet around
HD~38858, and then RV limits on undiscovered planets around HD~20794, 61~Vir (HD~115617),
HD~69830, and HD~38858. A detailed analysis of the High Accuracy Radial velocity Planet
Searcher \citep[HARPS,][]{2003Msngr.114...20M} results for these systems will be
presented in Marmier et al. (in preparation). We then go on to discuss possible relations
between the disc structure and known and unknown planets around these stars.

HD~38858b was announced by \citet{2011arXiv1109.2497M}, as a planet with a minimum mass
of 30.55 $M_\oplus$ orbiting at 1 au, using 52 RV measurements taken over eight
years. With additional data it has become clear that this signal was spurious and arises
due to an alias of the magnetic cycle of the star, which has a period of 2930 days. The
additional data (now 96 measurements with several years more coverage) have however
revealed a planetary signal with a period of $198 \pm 1$ days (0.64 au) and minimum mass
of $12 \pm 2 M_\oplus$.

\subsection{Limits on undiscovered planets}\label{ss:pl}

To consider the structure of the planetary systems in light of our results, we now
present the limits on undiscovered planets set by HARPS data. The limits depend primarily
on the RV precision ($K$ in m s$^{-1}$) and the time span over which the observations
were made. The limits are therefore approximately $M_{\rm pl} \sin i > 11 K \sqrt{M_\star
  a_{\rm pl}}$ in units of Earth masses (with $M_\star$ in units of $M_\odot$ and $a_{\rm
  pl}$ in au) for orbital periods less than the span of observations, and increase very
steeply for longer orbital periods (roughly as $a_{\rm pl}^{3.5}$).

\begin{figure}
  \begin{center}
    \hspace{-0.25cm} \includegraphics[width=0.48\textwidth]{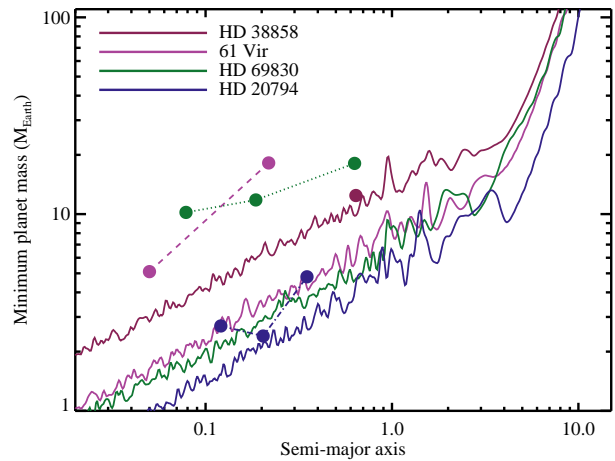}
    \caption{Planet minimum masses and semi-major axes for HD~20794, HD~69830, 61~Vir,
      and HD~38858 (dots, joined by dot-dashed, dashed, and dotted lines respectively,
      with no line for HD 38858b). Solid lines show the 3$\sigma$ limits on planets set
      by the HARPS RV data (the legend has the same vertical ordering as the
      lines).}\label{fig:rvlim}
  \end{center}
\end{figure}

Fig. \ref{fig:rvlim} shows the limits on unseen planets derived from an analysis of the
HARPS RV data. In each case the planetary signals (and stellar signals where necessary)
were first subtracted from the time series. Then, for each point in the mass-period
diagram, a corresponding circular orbit is added to the residuals, with 20 phase values
(spaced equidistantly between 0 and 2$\pi$). For each phase, this process is repeated
1000 times with the residuals randomly permuted, and we consider that the synthetic
planet is detected only if we see a signal in the resulting periodogram that is higher
than the 1 per cent false alarm probability limit at all phases.

All four systems have been observed for a time span of about 10 years, which is reflected
in the planet sensitivity in terms of semi-major axes. The differences in mass
sensitivity ($K$) are almost entirely due to differences in the number of measurements,
with 96 for HD~38858, 239 for HD~69830, 209 for 61~Vir, and 461 for
HD~20794. Fig. \ref{fig:rvlim} shows that planets interior to HD~38858b with masses
greater than about 10 $M_\oplus$ could have been detected in this system. Thus, while
61~Vir or HD~69830-like planetary systems can be ruled out around HD~38858, lower mass
planets (as seen around HD~20794) may still be present in this system. In all cases
Saturn-mass planets within about 10 au could have been detected.

\subsection{Radial structure and history}

\begin{figure}
  \begin{center}
    \hspace{-0.25cm} \includegraphics[width=0.48\textwidth]{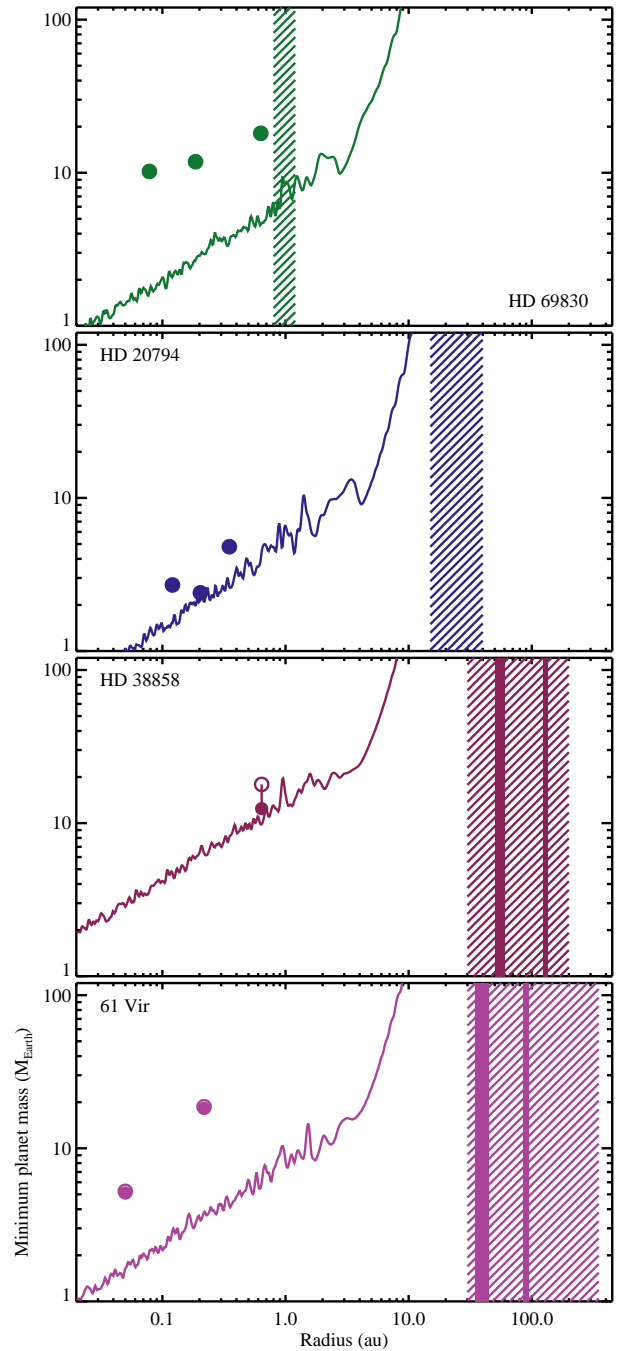}
    \caption{Cartoon depictions of the four nearby G-type systems with low-mass planets
      and debris disc detections. Increased planet masses around HD~38858 and 61~Vir
      assuming coplanarity with the resolved disc are included (open circles), but not
      visible for 61~Vir due to the small correction. RV limits from Fig. \ref{fig:rvlim}
      are shown as solid lines. The regions occupied by the discs are shown to the
      right. The uncertainty in the structure is indicated for HD~38858 and 61~Vir by
      including the two-belt model (solid region) and the extended models (hatched
      region). For HD~69830 and HD~20794 the approximate location is indicated by the
      hatched region.}\label{fig:sys}
  \end{center}
\end{figure}

Combining the RV limits, our above disc modelling results, those for 61~Vir from
\citet{2012MNRAS.424.1206W}, and those for HD~69830 from \citet{2007ApJ...658..584L},
Fig.~\ref{fig:sys} shows cartoon depictions of the radial structure in each system. As in
Fig. \ref{fig:rvlim}, planet masses and the limits on undetected planets are shown. The
disc extents are derived from the \emph{Herschel} images, or constrained by mid-IR
imaging and modelling of the disc spectrum in the case of HD~69830
\citep{2007ApJ...658..584L,2009A&A...503..265S}. The hatched regions cover the
approximate radial locations where the dust is seen, with the solid regions for 61~Vir
and HD~38858 showing the location of the two-belt models. We have not included the hot
dust seen around HD~20794 \citep{2014A&A...570A.128E} as the origin of this phenomenon is
unknown, and may be more related to stellar physics than planetary systems.

Contrasting the four systems, most notable is that HD~69830 does not possess a
significant level of cool dust \citep{2014A&A...565A..15M}, suggesting that this system
evolved in a different way to the other three, and that the dust at 1 au is perhaps
short-lived and directly linked to one or more of the three planets. This system has been
discussed extensively elsewhere
\citep[e.g.][]{2005ApJ...626.1061B,2006A&A...455L..25A,2007ApJ...658..584L,2007ApJ...658..569W,2009MNRAS.393.1219P,2009A&A...503..265S,2011ApJ...743...85B},
so here we focus on the remaining three.

Our main goal is to explore links between the planets and the disc structure, whether
these links provide potentially useful information about the system histories, and which
observations are needed to distinguish among different possibilities. We first highlight
some relevant similarities and differences among these three systems: The discs around
HD~20794, 61~Vir, and HD~38858 have similar inner radii of around 20-30 au, and (assuming
that the minimum masses are similar to the true planet masses) the planets around 61~Vir
and HD~38858 are more massive than those around HD~20794. The discs around 61~Vir and
HD~38858 extend to large radii ($>$100 au), and while the constraints are much poorer,
the above modelling shows that the HD~20794 disc is at least relatively depleted outside
$\sim$30 au. From the SEDs and images of 61~Vir and HD~38858 we can also be confident
that the regions inside $\sim$30 au are significantly depleted of debris compared to the
levels seen between 30-200 au. Being very faint and marginally resolved at best, the
relative constraints for HD~20794 on the regions inside the detected belt at $\sim$20 au
are poor.

Ideally we would like to link the properties of the known planets with the observed disc
structure, thus avoiding the invocation of unseen planets with uncertain
properties. Secular perturbations provide one way that planets can reach over large
radial distances and affect planetesimal belts
\citep[e.g.][]{2009MNRAS.399.1403M}. However, the secular perturbation timescale for the
outermost planet to affect the disc inner edge is $>$10 Gyr for all three systems, so
such a link is not possible given the current orbits of the known planets.

In the absence of strong effects from secular perturbations, we can still make a
speculative link between the masses of the known planets and the radial width of the
discs, which are larger for the two discs in systems with more massive
planets. Super-Earths may have difficulty forming in situ
\citep[e.g.][]{2005ApJ...631L..85Z,2006A&A...455L..25A,2014MNRAS.440L..11R,2014arXiv1412.4440I},
so could have formed farther out where the discs are observed (i.e. near 20 au),
imprinted some structure on the disc, and then migrated inwards to their observed
locations.
% With only three systems this link is of course speculative, but is worth discussing to
% motivate further theoretical and observational work. It has been argued that super-Earth
% mass planets do not form in situ, and therefore that they or the embryos from which they
% formed migrated inwards from more distant formation regions
% \citep[e.g.][]{2005ApJ...631L..85Z,2006A&A...455L..25A,2014MNRAS.440L..11R}. Thus, a
% plausible link can be made in a scenario where the observed planets formed farther out
% where the discs are observed (i.e. near 20 au) , imprinted some structure on the disc,
% and then migrated inwards to their observed locations.
In this kind of scenario the planet mass dependence of the disc radial width arises
because more massive planets excite planetesimals to greater eccentricities. That is, at
least the outer planets around HD~20794, 61~Vir, and HD~38858 all formed near the
observed inner disc edges, but the $>$10 $M_\oplus$ outer planets around 61~Vir and
HD~38858 were better able to scatter planetesimals onto eccentric orbits compared to the
$<$10 $M_\oplus$ planets around HD~20794.
% because the velocity ``kick'' imparted by a planet in a close encounter with a
% planetesimal is roughly the escape velocity from the planet ($v_{\rm esc}$), and hence
% larger for more massive planets. The maximum eccentricity of the planetesimals depends
% on this velocity relative to the local Keplerian velocity, $v_{\rm esc}/v_{\rm K}$ (or
% relative to the escape velocity from the star). This relation is commonly used as a
% planetesimal ejection criterion, but in a planet-forming disc the effects of gas drag
% will damp eccentricities, meaning that the kicks are less effective and that ejection
% is harder. The qualitative result is therefore that at a given stellocentric distance,
% more massive planets excite planetesimals onto more eccentric orbits.  Applying this
% idea here, a possible explanation is that the extent of the discs around 61~Vir and
% HD~38858 is greater than for HD~20794 because their planetesimals were more strongly
% scattered by the more massive planets.
The depletion of disc material between the current planet and disc locations is explained
in this scenario by i) the use of this material to form the observed planets, and/or ii)
the shepherding of this material inwards as the super-Earths migrated towards the star
\citep[e.g.][]{2005ApJ...631L..85Z,2008ApJ...682.1264K,2014ApJ...794...11I}.

By analogy with the Solar System, the radial width of the discs around 61~Vir and
HD~38858 would therefore be interpreted as arising from ``scattered disc''-like
populations of planetesimals on eccentric orbits. A prediction of this scenario is that
the radial width of the disc as observed in small dust with \emph{Herschel} is a property
inherited from the larger planetesimals, and therefore that the disc extent should also
appear to be wide when larger grains that are less affected by radiation forces are
imaged at mm wavelengths. This appearance would be in contrast to discs such as $\beta$
Pic and AU~Mic, where the planetesimals occupy a relatively narrow ``birth ring'' and the
small grains that are strongly affected by radiation forces reside in a wider halo
\citep{2006ApJ...648..652S}.

% A more general prediction of this a scenario is that the width of the disc should scale
% with planet mass (i.e. $\Delta r_{\rm disc}/r_{\rm disc}$ scales with $v_{\rm esc}/v_{\rm
%   K}$). Given that planets of different masses migrate in different ways under different
% circumstances \citep[e.g.][]{2007prpl.conf..655P,2007prpl.conf..669L}, such a relation is
% unlikely to be universal, but provides a simple hypothesis for interpreting systems with
% planets and debris discs, and seems most applicable in the case of low-mass planets due
% to their lesser ability to eject planetesimals in close encounters.

While this scenario is convenient in that it does not require undetected planets at
larger semi-major axes, it is by no means unique and the sensitivity limits in
Figs. \ref{fig:rvlim} and \ref{fig:sys} show that the constraints allow $\gtrsim$10-20
$M_\oplus$ planets beyond about 5 au, and $>$Saturn-mass planets beyond 10
au.\footnote{Being old stars, these systems are not high-priority targets for
  high-contrast direct imaging surveys because the sensitivity to $\sim$Gyr-old giant
  planets is poor; only HD~69830 has been observed recently, and was targeted based on
  the presence of the disc \citep{2013ApJ...773...73J}.} Given the total mass of solids
in a ``typical'' planet-forming disc is roughly 10-100 $M_\oplus$, and the gas mass
$\sim$100 times greater \citep[e.g.][]{1977Ap&SS..51..153W,2005ApJ...631.1134A}, the
existence of more planets is certainly possible given the mass budget. Another possible
scenario is therefore that the known planets formed closer to the star, or at least from
material that originated closer to the star \citep[e.g.][]{2012ApJ...751..158H}, and that
extra planets may (or may not) exist in the regions where the RV limits are poor. In this
case any links between the masses of the known super-Earths and the disc structure would
be far more tenuous, and likely non-existent. An obvious question is then whether
undetectable planets are responsible for the different radial widths and the disc
depletion inside 20 au, and which signatures can distinguish planet formation and/or
planet clearing from other depletion scenarios.

Firstly, planets are not necessarily needed to clear the inner disc regions. Most simply
it may be that planetesimal formation was less efficient in the 5-20 au region than
elsewhere and that these regions are naturally empty.
% The tentative detection of CO described in section \ref{sss:co} highlights the
% possibility of release of this gas from planetesimals that formed beyond the CO
% ``snowline'', and it may be that planetesimal formation is most efficient at such
% locations \citep[previously being applied to the water,
% e.g.][]{1988Icar...75..146S,2004ApJ...614..490C}.
In the event that planetesimals did form in this region, \citet{2012MNRAS.424.1206W} show
that the disc around 61~Vir may simply be collisionally depleted from the inside out, and
given the uncertain stellar ages, the same argument can be applied to HD~20794 and
HD~38858. This scenario predicts a characteristic slope for the brightness of the disc
inner edge \citep{2008ARA&A..46..339W,2010MNRAS.405.1253K}. The main prediction if the
inner disc edges are instead truncated by unseen planets is that the edge should be
sharper than for collisional depletion \citep{2006MNRAS.372L..14Q,2012MNRAS.419.3074M},
and that any relation between the known planet masses and the disc extent is
coincidental. The best way to tell how the inner disc regions were depleted is therefore
to resolve the inner disc edge, a task best suited to high resolution optical or
interferometric mm imaging \citep[e.g.][]{2005Natur.435.1067K,2012ApJ...750L..21B}.  The
radial disc extent may of course also be related to the masses of these unseen planets,
but unless these planets are detected or inferred to exist there is little power to tell
whether their influence on the width was significant. However, their possible presence
highlights other scenarios. The scattered disc component of the Kuiper belt formed as
planetesimals were scattered by the outward migration of Neptune
\citep{1997Sci...276.1670D}, providing an alternative explanation for the wider extent of
the discs around 61~Vir and HD~38858 noted above, but no reason (aside from lower planet
masses or the lack of such a process) for the narrower belt around HD~20794. The
additional planets required in this scenario would be very difficult to detect, so a key
signature that would distinguish among these scenarios is that outward migration captures
planetesimals in mean motion resonances \citep{2003ApJ...598.1321W}, with the resultant
clumpy structures expected to be detectable in high resolution sub-mm images
\citep{2012A&A...544A..61E}.

In summary, the greater disc radial widths and planet masses around 61~Vir and HD~38858
compared to HD~20794 are suggestive of a link between the two; the wider extent may be a
signature of a scattered disc population of planetesimals that was excited by the planets
before they migrated to their current locations. Being based on three systems this
scenario is clearly speculative and other scenarios are of course also possible, so such
ideas provide general motivation for further characterisation of these and other
systems. In particular, high resolution observations that measure the shape of the inner
disc edges and any azimuthal asymmetries will provide the best constraints on the
structure and origins of these systems.

\section{Conclusions}\label{s:conc}

We present \emph{Herschel} and JCMT observations of the debris discs in the Sun-like
planet hosting systems HD~38858 and HD~20794. The HD~38858 disc is well resolved and
spans a wide range of radii, from approximately 30-50 au to 130-200 au depending on
whether the dust lies in two narrow rings or a wide continuous belt. We present a
probable 850 $\mu$m imaging detection of the disc and set an upper limit on the level of
CO J=2-1 emission. The HD~20794 disc appears robust photometrically, but is marginally
resolved at best. The disc has a probable radius of a few tens of au and is among the
faintest known, with dust levels roughly one order of magnitude above the
Edgeworth-Kuiper belt in the Solar System \citep[e.g.][]{2012A&A...540A..30V}. Using the
non-detection of excess emission at 24 $\mu$m we show that the hot dust seen with near-IR
interferometry probably lies interior to the innermost planet at 0.1 au.

With the addition of 61~Vir and HD~69830, we consider the sample of four nearby G-type
stars with low-mass planets and debris disc detections. We present new limits on
undiscovered planets in these systems using HARPS radial velocity data, showing that
regions beyond about 5 au may contain undiscovered $\gtrsim$30 $M_\oplus$ planets. With
the exception of HD~69830, the detected disc components lie well beyond the known planets
in these systems and various different scenarios can explain the observed system
structure. A scenario that does not invoke additional planets is that the extent of the
observed discs around HD~20794, 61~Vir, and HD~38858 is related to the mass of the
planets, which are greater around the latter two stars. Other scenarios are possible
however, and high resolution imaging that can determine the shape of the disc inner
edges, and the extent of the parent planetesimals, will help distinguish among them.

\section*{Acknowledgments}

We thank the referee for a thoughtful review. This work was supported by the European
Union through ERC grant number 279973 (GMK, LM, \& MCW). LM also acknowledges support by
both STFC and ESO through graduate studentships. MM, CL, FP, \& SU acknowledge the Swiss
National Science Foundation (SNSF) for the continuous support of the radial-velocity
research programmes.

% \bibliography{../ref} \bibliographystyle{apj}

\end{document}